\documentclass[accept]{elsarticle}

\usepackage{hyperref}

\usepackage[normalem]{ulem}

\usepackage{subfigure}
\usepackage{graphicx}
\usepackage{amssymb,amsfonts}
\usepackage{amssymb,amsfonts,amsmath,latexsym,dsfont}
\usepackage[ruled]{algorithm2e}
\usepackage{amsthm}
\usepackage{mathtools}
\usepackage{wasysym}
\usepackage{wrapfig}

\usepackage{tikz,pgfplots}
\usetikzlibrary{plotmarks}

\usepackage{fullpage}
\usepackage{graphicx}
\usepackage{mathrsfs}
\usepackage{framed,multirow}
\usepackage{setspace}

\usepackage{booktabs}

\usepackage{soul}

\newdimen\iwidth
\newdimen\iheight

\newcommand{\Tiny}[1]{\scalebox{0.6}{#1}}

\newcommand{\mc}[3]{\multicolumn{#1}{#2}{#3}}

\graphicspath{{Fig/}{./}}

\newcommand{\curl}{\ensuremath{\nabla\times\,}}

\renewcommand{\div}{\nabla\cdot}
\newcommand{\grad}{\nabla}

\newcommand{\gradh}{\ensuremath{\nabla}_h}

\newcommand{\im}{\textit{\i}}

\newcommand{\bfe}{{\bf e}}

\newcommand{\bfp}{{\bf p}}
\newcommand{\bfx}{ {\bf x}}

\newcommand{\bfu}{{\bf u}}
\newcommand{\bfq}{{\bf q}}

\newcommand{\bfr}{{\bf r}}

\newcommand{\bfgamma}{{\boldsymbol \gamma}}

\newcommand{\bfmu}{{\boldsymbol \mu}}
\newcommand{\bflambda}{{\boldsymbol \lambda}}
\newcommand{\bfrho}{{\boldsymbol \rho}}
\newcommand{\bfkappa}{{\boldsymbol \kappa}}

\newtheorem{definition}{Definition}[section]

\journal{Journal of Computational Physics}

\begin{document}

\begin{frontmatter}

\title{A hybrid shifted Laplacian multigrid and domain decomposition preconditioner for the elastic Helmholtz equations}
\author[mymainaddress]{Eran Treister\corref{mycorrespondingauthor}}\ead{erant@cs.bgu.ac.il}
\cortext[mycorrespondingauthor]{Corresponding author}
\author[mymainaddress]{Rachel Yovel}\ead{yovelr@bgu.ac.il}

\address[mymainaddress]{Department of Computer Sciences, Ben-Gurion University of the Negev, Beer Sheva, Israel.}
\tnotetext[mytitlenote]{This research was supported by The Israel Science Foundation (grant No. 1589/19). The research has also received funding from the European Union's Seventh Framework Programme (FP7/2007-2013) under grant agreement number 623212---MC Multiscale Inversion. The authors also thank the Lynn and William Frankel Center for Computer Science at BGU.}

\begin{abstract}
In this work we extend the shifted Laplacian approach to the elastic Helmholtz equation. The shifted Laplacian multigrid method is a common preconditioning approach for the discretized acoustic Helmholtz equation. In some cases, like geophysical seismic imaging, one needs to consider the elastic Helmholtz equation, which is harder to solve: it is three times larger and contains a nullity-rich grad-div term. These properties make the solution of the equation more difficult for multigrid solvers. The key idea in this work is combining the shifted Laplacian with approaches for linear elasticity. We provide local Fourier analysis and numerical evidence that the convergence rate of our method is independent of the Poisson's ratio. Moreover, to better handle the problem size, we complement our multigrid method with the domain decomposition approach, which works in synergy with the local nature of the shifted Laplacian, so we enjoy the advantages of both methods without sacrificing performance. We demonstrate the efficiency of our solver on 2D and 3D problems in heterogeneous media.
\end{abstract}


\end{frontmatter}


\section{Introduction}
The Helmholtz equation is used to model the propagation of a wave within a heterogeneous
medium. Its acoustic version is given by
\begin{eqnarray}
\label{eq:acousitcHelm}
\rho\; \div \left(\rho^{-1}\grad p\right) + \omega^{2}\kappa^2p = q,
\end{eqnarray}
where $p = p(\vec{x}), \vec{x}\in\Omega$ is the Fourier transform of the wave's \emph{pressure} field, $\omega = 2\pi f$ is the angular frequency, $\kappa = \kappa(\vec{x}) > 0$
is the ``slowness'' of the wave in the medium (the inverse of the wave velocity), and $\rho = \rho(\vec{x}) > 0$ is the density of the medium. The right-hand-side $q(\vec{x})$ incorporates sources into the equation. The equation is discretized on a finite domain and is accompanied with some absorbing boundary conditions (ABC) \cite{engquist1977absorbing}, that mimic the propagation of a wave in an open domain. This is usually achieved by some complex-valued absorbing boundary layer \cite{engquist1979radiation,singer2004perfectly}, which is related to modeling the attenuation of the wavefield. 

The acoustic equation \eqref{eq:acousitcHelm} is usually discretized by a finite-difference scheme on a regular grid, resulting in a large and indefinite linear system, which is complex-valued due to the absorbing boundary conditions and possible attenuation. If the frequency $\omega$ (or the wavenumber $\kappa\omega$) is high, the problem requires a very fine mesh and a large number of unknowns \cite{bayliss1985accuracy,haber2011fast}. In this case, solving the discretized equation at large scale 3D scenarios is challenging, and is still considered to be an open problem.

One of the main applications that include the Helmholtz equation is full waveform inversion \cite{pratt1999, virieux2009overview, metivier2017full, JointEikFWI17,eslaminia2022full}, which is a process used to estimate the wave velocity and rock structure of the earth's subsurface.
The inversion process (in the frequency domain) includes many repeated solutions of Helmholtz equations for modeling the wave propagation. These solutions are used to iteratively estimate the unknown wave velocity in the earth's subsurface.
However, because the earth is an elastic medium, the acoustic equation in \eqref{eq:acousitcHelm} does not fully capture the physics of the wave propagation, and research is advancing towards elastic waveform inversion, in which the elastic Helmholtz equation is solved for modeling the wave propagation \cite{brossier2009seismic, brossier2011two,borisov2015three, kormann2017acceleration}.

The elastic Helmholtz equation, which we formulate later, is a system of partial differential equations (PDEs). While \eqref{eq:acousitcHelm} models pressure waves only, the elastic Helmholtz equation also models shear waves. Similarly to \eqref{eq:acousitcHelm}, the linear system that results from discretizing the elastic equation is indefinite and complex-valued. Moreover, because the equation is a system of PDEs, the associated linear system is three times larger (in 3D) than the acoustic one for the same mesh size. In addition, the discretization requires the mesh to be finer than in the acoustic \eqref{eq:acousitcHelm}, because the modeled shear waves have higher wavenumber than the pressure waves \cite{martin2006marmousi2}. Altogether, we get a huge linear system which is more difficult to solve than the acoustic one and an iterative method is required for its solution. However, while the solution of the acoustic equation \eqref{eq:acousitcHelm} has been heavily studied in the literature with a variety of methods \cite{poulson2013parallel,haber2011fast,livshits2014scalable,olson2010smoothed,gordon2013robust,Treister2018point,stolk2013rapidly,gander2013domain, chen2016robust,cosnabrugge2016convergent,wang2020taylor}, 
the elastic version has very few available iterative solvers known to us. One recent ``elastic solver'' is \cite{li20152d}, which is an extension of \cite{gordon2013robust} to the elastic case. This method, which involves a hybrid parallel Kaczmarz preconditioner, is quite generic, and hence requires many iterations to solve the system at large scales.

One of the most common solvers for the discretized acoustic equation \eqref{eq:acousitcHelm} is the shifted Laplacian multigrid method  \cite{erlangga2006novel,umetani2009multigrid,oosterlee2010shifted,tsuji2015augmented,calandra2013improved,cools2014new,cools2015multi,Tobias2017}, 
where an attenuated version of \eqref{eq:acousitcHelm} is used as a preconditioner for the original system inside a Krylov method. The attenuated system, which is the same system with a complex shift, can be easily solved by multigrid, if the attenuation is high enough. However, as we add more attenuation, the efficiency of preconditioner deteriorates. This is a tradeoff that methods try to balance.

The shifted Laplacian multigrid approach seems to be naturally extendable to the elastic case. Indeed, the work of \cite{airaksinen2009damping} applies the standard shifted Laplacian method for the problem using algebraic multigrid operators. However, experiments show that the standard shifted Laplacian multigrid is not efficient for the elastic Helmholtz equation, and some specialized treatment is necessary. The recent \cite{rizzuti2016multigrid} suggests such a multigrid method, using line-relaxation instead of point-wise relaxation. However, this method is only presented for 2D problems and does not seem to achieve the same efficiency compared to the acoustic case. In particular, the authors use a significantly higher attenuation (shift) parameter than what is usually used in the acoustic case, and the line relaxations in 2D extend to relatively expensive plane relaxations in 3D. These two properties leave room for improvement.

\paragraph{Contribution}

Our main contribution in this work is the development of a new shifted Laplacian multigrid method for the elastic Helmholtz equation. 
As far as we know, we are the first to suggest a shifted Laplacian multigrid method for the elastic Helmholtz equation, with performance comparable to the well-studied shifted Laplacian method for the acoustic equation. Our methods scales well for the nearly incompressible case, as we demonstrate both in our numerical results and in our theoretical local Fourier analysis. We further improve the suggested multigrid method by combining it with domain decomposition, to enhance parallelism and deal with the size of the problem. We observe that the local nature of both methods enables this combination without a significant loss in convergence rate. The method is shown to tackle 2D and 3D cases with challenging heterogeneous velocity models.

Our multigrid method adopts approaches that are suitable for linear elasticity to better treat the ``elastic part'' of the elastic Helmholtz equation. This part includes the weighted Laplacian (as in \eqref{eq:acousitcHelm}), with an additional grad-div term that has a rich null-space. This part is essentially the elasticity operator and is known to cause difficulties to standard multigrid methods in cases of nearly incompressible material. The shifted Laplacian method is no exception, and to solve the elastic equation using multigrid, we suggest applying the mechanisms for both the elasticity and the acoustic Helmholtz equation together. To this end, we write the elastic equation using a mixed formulation \cite{gaspar2008distributive} and use a local cell-wise ``Vanka'' relaxation to treat the elastic part of the elastic Helmholtz equation \cite{wobker2009numerical}. The indefiniteness of the problem is treated by shifted Laplacian in the same way that the indefiniteness of \eqref{eq:acousitcHelm} is treated. We demonstrate that our method is scalable with respect to the Poisson ratio: it performs similarly to standard shifted Laplacian for \eqref{eq:acousitcHelm}, regardless of the Poisson ratio, only with respect to the shear wavenumber instead of the pressure wavenumber.

The only pitfall with our multigrid method, preventing it from being applicable for large 3D cases, is the memory consumption. Multigrid is also cumbersome to parallelize across multiple machines. To this end, in our second contribution we explore the combination of the shifted Laplacian multigrid and the domain decomposition (DD) iterative methods \cite{benamou1997domain, bank2013domain, stolk2013rapidly,gander2013domain, chen2016robust}. These two are among the most common methods for solving the acoustic equation. DD approaches involve decomposing the problem into subregions, solving each subproblem separately, and attaching the local solutions together.
This procedure is repeated iteratively as a preconditioner in a Krylov method, and the sub-domain division yields a natural and easy way to distribute or parallelize the solution of the preconditioned system across several workers or computing nodes. The solution of each sub-domain problem is usually achieved by a direct solver. However, DD methods generally preform better when one uses less subdomains or increases overlap between the subdomains. Both yield larger local problems which for our case are expensive to solve using an LU factorization. On the other hand, if the domain is divided aggressively, convergence is hampered.
We propose to use moderate subdomain sizes and solve them with multigrid so that the memory footprint is low. 

\begin{figure}
\begin{center}
	\newcommand{\image}[1]{\includegraphics[width=0.32\linewidth]{./images/#1}}
    \subfigure[\footnotesize Standard solution]{\image{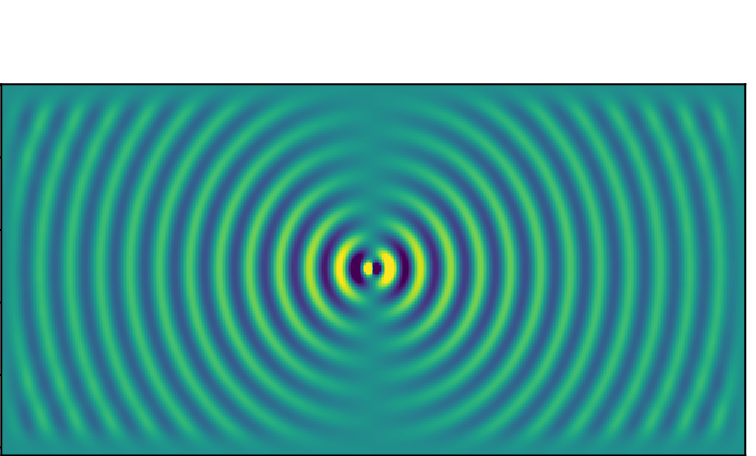}}\hspace{80pt}
    \subfigure[\footnotesize Attenuated solution]{\image{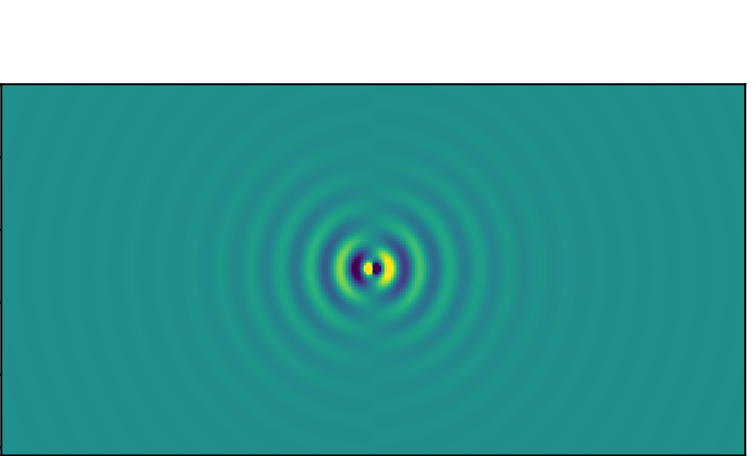}}\\
\end{center}
\caption{\footnotesize The $u_x$ component of the elastic Helmholtz solution for a point source in a constant medium, in a standard and attenuated  scenarios (with attenuation of $0.15$). The attenuated solution approximates the standard solution only in the local region near the point source.}
\label{fig:shifted}
\end{figure}

Our main observation here is that when using shifted Laplacian solvers, it is not necessary to capture global information. The added attenuation yields solution with a local support only, (see Fig. \ref{fig:shifted}), hence, dividing the domain into subdomains and solving the problem locally will not harm the convergence of shifted Laplacian. In other words, we use the added value of domain decomposition to exploit the inevitable locality weakness generated by the added attenuation in shifted Laplacian. In this way we enjoy the advantages of both methods, and we show that the combination can yield the same convergence properties of each of the methods alone.

Our paper is organized as follows: in Section \ref{sec:Backg} we give mathematical preliminaries: we present the elastic Helmholtz equation, discuss its discretization and briefly present the general shifted Laplacian method and the DD method. In Section \ref{sec:Method} we present our multigrid method, and discuss the relation between the acoustic and elastic equations in the case of fully incompressible material. In Section~\ref{sec:hybridDD} we present the combination of the DD method into our hybrid preconditioner. In Section \ref{sec:LFA} we hold local Fourier analysis for the Vanka smoother we use in the multigrid cycle. Finally in Section \ref{sec:NumericalResults} we demonstrate the properties and efficiency of our method in a few numerical examples in two and three dimensions.

\section{Mathematical background} \label{sec:Backg}
\subsection{Problem formulation and discretization}
The elastic Helmholtz equation has several formulations. Here we focus on the equation in isotropic medium, which is formulated by either of the following equivalent\footnote{This equivalence holds for constant coefficients. However, the second formulation can be used as a preconditioner for the first one in heterogeneous cases.} equations:
\begin{eqnarray}\label{eq:elasticHelm}
 \grad\lambda\div\vec u + \div\mu\left(\grad\vec u+\grad\vec u^T\right) +\omega^{2}\rho\vec u &=& \vec q_{s}, \quad \text{or}\\
\nonumber\grad(\lambda + \mu)\div\vec u + \div\mu\grad\vec u +\omega^{2}\rho\vec u &=& \vec q_{s}.
\end{eqnarray}
The unknown $\vec u = \vec u(\vec x)$ is a displacement vector which, in three dimensions, has  three components at each location in the domain. $\mu = \mu(\vec x)$ and $\lambda = \lambda(\vec x)$ are the Lam{\'e} parameters,  and $\rho$ is the density of the medium as in \eqref{eq:acousitcHelm}. These parameters determine the pressure and shear wave velocities by $V_p = \sqrt{(\lambda+2\mu)/\rho}$, and $V_s = \sqrt{\mu/\rho}$, respectively \cite{li2016Fourth}.
The term $\div \mu\grad\vec u$ is the weighted diffusion operator $\div\mu\grad$ applied on each of the components of the vector $\vec u$ separately. In the case of $\omega = 0$, \eqref{eq:elasticHelm} becomes the elasticity operator. In the case where $\mu = 0$ the material is incompressible. Then, as we show in Section \ref{sec:Method}, the elastic equation can be reduced to the acoustic equation \eqref{eq:acousitcHelm}, modeling only pressure waves, with $V_p$ as the pressure wave velocity. Other richer elastic formulations may include more parameters than $\lambda$ and $\mu$, e.g. the orthorhombic formulation has 9 parameters and also models anisotropy \cite{li20152d}. In principle, the method that we present in this paper is suitable for anisotropic cases as well, as long as the anisotropy is not too strong. If the anisotropy is strong, our method will require adaptations similar to those that are needed for the standard shifted Laplacian method for an anisotropic Laplacian operator in \eqref{eq:acousitcHelm}, e.g. semi-coarsening. Such extensions are beyond the scope of this paper.

To discretize \eqref{eq:elasticHelm} using a finite-differences scheme on a regular mesh, one have to choose between two approaches: node-based or staggered grid discretization. In the former, the displacement components $u_1,u_2$ and $u_3$ are located at the nodes of a grid cell, and in the latter the displacement components are located on the faces of the cell. The advantage of the nodal approach is the ability to formulate high-order or optimally weighted second-order discretizations using a compact 27-point stencil \cite{kelly1976synthetic,vstekl1998accurate,gosselin20143d} at each component. Such stencils allow fast memory access in matrix-vector products, and low fill-in when using direct solvers such as MUMPS \cite{MUMPS2001} or PARDISO \cite{schenk2004solving}. For this reason, it is a common approach for discretizing the acoustic equation \eqref{eq:acousitcHelm}, see \cite{operto20073d, singer1998high, singer2006sixth, turkel2013compact}. However, in the case of the elastic equation \eqref{eq:elasticHelm} for nearly incompressible materials when $\mu \ll \lambda$, the nodal approach leads to relatively large errors \cite{rizzuti2016multigrid}, whereas the staggered discretization is stable for nearly incompressible materials \cite{virieux1986p,levander1988fourth}. On the other hand, compact high-order staggered discretizations are currently not available. For example, the fourth order schemes of \cite{levander1988fourth,li2016Fourth} create stencils that are wider than the 27-point stencil block to minimize dispersion errors. This, however, leads to relatively high memory access time in matrix-vector products, and high fill-in using direct methods.

\begin{figure}
  \centering
  \includegraphics[scale=0.12]{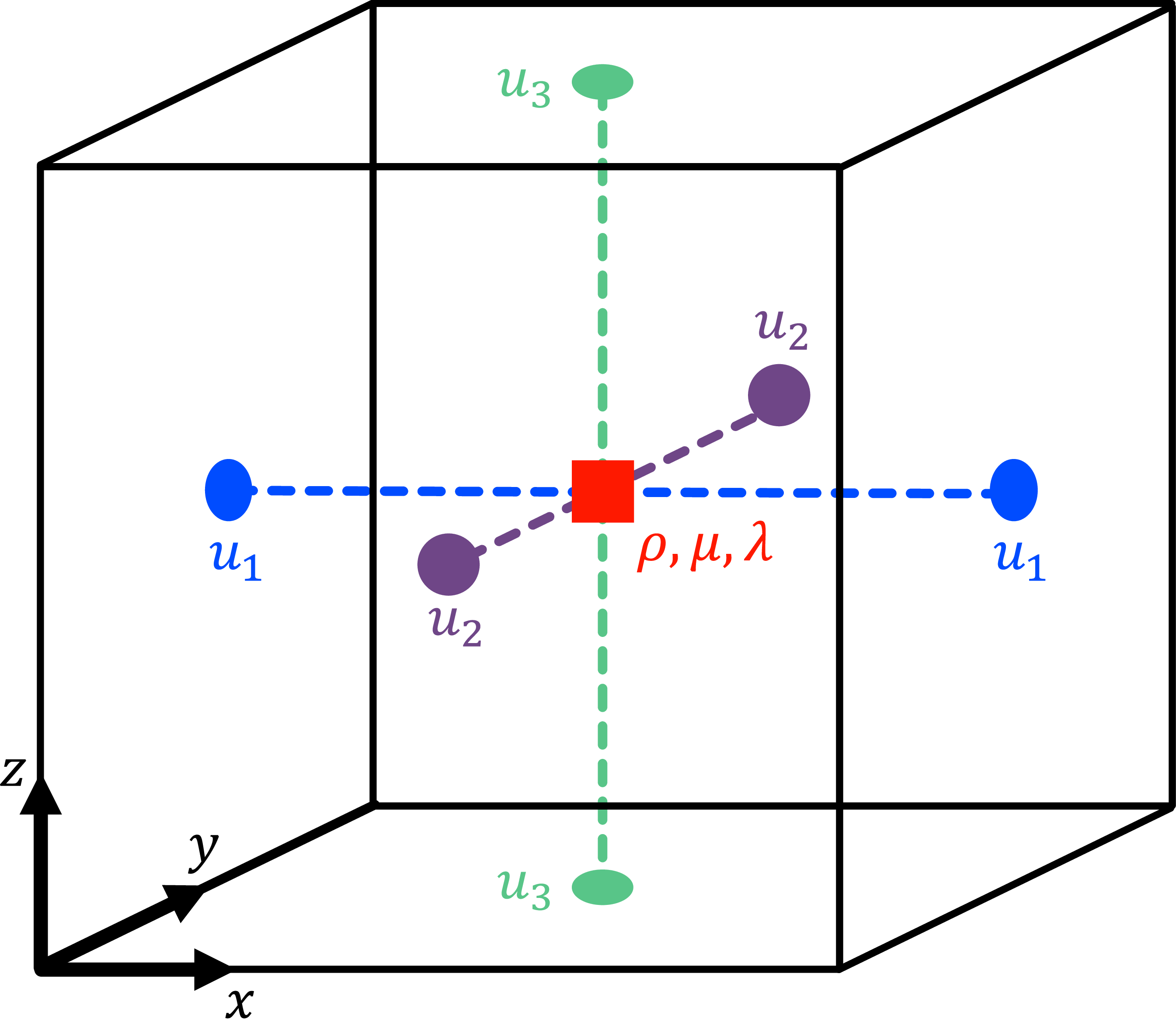}\\
  \caption{The staggered grid discretization of a cell in 3D.}\label{fig:stag}
\end{figure}

In this work we focus on a multigrid solver for \eqref{eq:elasticHelm}, and use a standard second-order staggered grid discretization, which is illustrated in Fig. \ref{fig:stag}. This discretization was also used in \cite{virieux1986p,rizzuti2016multigrid}, and suggested in \cite{TOS01} to discretize similar systems of PDEs.
We place the components of $\vec u$ on the faces of the cell, and denote its discrete components in boldface, i.e., the vector $\vec\bfu$ has the discrete values of $\vec u$ on the mesh. We place $\rho,\mu,\lambda$ at the cell center and similarly denote their discrete vectors by $\bfrho,\bfmu,\bflambda$.

We now obtain the linear system that results from the discretization of the bottom formulation in~\eqref{eq:elasticHelm}. To disctretize the first derivatives, we use the second order central difference scheme
\begin{equation}
\frac{\partial v}{\partial x}(x) \approx \frac{v(x+h/2) - v(x-h/2)}{h},
\end{equation}
which is used for the gradient and divergence operators. The resulting linear system is given by
\begin{equation}\label{eq:elasticHelmSystem}
H^e\vec{\bfu} = \left(\gradh D_c(\bflambda+\bfmu)\gradh^T + \vec{\grad}_h^TA_e(\bfmu)\vec{\grad}_h - \omega^2M\right)\vec{\bfu}  =  \vec{\bfq},
\end{equation}
where $\gradh$ is the cell-centered gradient operator that operates from the cell centers to the faces, at which the components $\bfu_i$ are placed. The matrix $-\gradh^T$ is used for the discrete divergence operator. The operator $D_c(\bflambda+\bfmu) = \mbox{diag}(\bflambda+\bfmu)$ generates a diagonal matrix with the values of $\bfmu+\bflambda$ on the cell centers. The operator $\vec{\grad}_h$ is a block diagonal gradient matrix, which includes three gradient matrices on its diagonal---one for each of the components $\bfu_1,\bfu_2,\bfu_3$---using central difference schemes. 
$A_e(\bfmu)$ averages the values of the cell-centered $\bfmu$ to the edges, and creates a diagonal matrix to hold the averaged values. Altogether, the term $\vec{\nabla}_h^TA_e(\bfmu)\vec{\grad}_h$ ends up being the weighted Laplacian operator which is applied on each of the components $\bfu_i$ separately. We define the mass matrix
\begin{equation}\label{eq:elasticHelmMass}
M = A_{f}(\bfrho\odot(1-\im\bfgamma/\omega)),
\end{equation}
where $A_{f}(\bfrho)$ is a diagonal matrix with the averaged values of a cell-centered $\bfrho$ onto the cell faces, and $\im$ stands for the imaginary unit. The symbol $\odot$ is the Hadamard product, and the vector $\bfgamma>0$ is a physical attenuation vector. We also use $\bfgamma$ to incorporate the absorbing boundary conditions, using a function that quadratically goes from zero to one towards the domain boundaries \cite{clayton1977absorbing}. The attenuation can be equivalently modeled by using complex frequency-dependent Lam{\'e} parameters \cite{vstekl1998accurate}, which can also be used to model different attenuation factors for the shear and pressure wave velocities \cite{gosselin20143d}. We place $\bfgamma$ at the cell center and assume that the physical attenuation is very small. Another approach for the absorbing boundary conditions may be the perfectly matched layer in \cite{collino2001application}.

One nice feature of the staggered discretization is that both the equations in \eqref{eq:elasticHelm} are equal \emph{even after the discretization} (for constant coefficients and up to boundary conditions). That is, we get two identical linear systems for each one of the formulations, up to small differences that result from averaging the coefficients and boundary conditions only. 
Finally, it is worthy to note that we use this second order scheme, as in \cite{virieux1986p,rizzuti2016multigrid}, to demonstrate our solver, but expect that it will also be effective for the high-order staggered discretizations \cite{levander1988fourth,li2016Fourth}. That is, we expect that, similarly to \cite{umetani2009multigrid}, the shifted Laplacian methods will work with similar efficiency for both second order and high-order nodal schemes for the acoustic case.

\subsection{Multigrid methods and the shifted Laplacian framework}
In this section we describe the general shifted Laplacian multigrid framework that we adopt in this paper. Multigrid methods aim at solving linear systems
\begin{equation}\label{eq:linsys}
H\bfu=\bfq
\end{equation}
iteratively by using two complementary processes. The first process is the relaxation, which is obtained by a standard local iterative method like Jacobi or Gauss-Seidel. Such methods are typically effective at reducing only part of the error in the iterative solution process. The other part of the error, called ``algebraically smooth'', is not reduced well by the relaxation and is typically defined by vectors $\bfe$ such that
\begin{equation}\label{eq:algSmooth}
\|H\bfe\| \ll \|H\|\|\bfe\|.
\end{equation}
To reduce these errors, multigrid methods use a ``coarse grid correction''. In this correction, the error $\bfe$ for some iterate $\bfx^{(k)}$ is estimated by solving a coarser system
$$
H_c\bfe_c = \bfr_c = P^T(\bfq-H\bfu^{(k)}),
$$
where the matrix $H_c$ approximates the matrix $H$ on a coarser mesh (the subscript $c$ denotes coarse components). The matrix $P$ is the so-called prolongation operator that is used to interpolate the solution of the coarse system, $\bfe_c$, back to the fine grid:
\begin{equation}
\bfe = P\bfe_c,
\end{equation}
and its transpose is used to restrict the residual onto the coarser grid.
The coarse operator $H_c$ can be obtained by either re-discretizing the problem on a coarser grid or by the Galerkin operator
\begin{equation}\label{eq:coarseGridMatrix}
H_{c} = P^T HP.
\end{equation}
Algorithm \ref{alg:TwoCycle} summarizes the process using two grids. By treating the coarse problem recursively, we obtain the multigrid V-cycle, and by treating the coarse problem recursively twice (by two recursive calls) we obtain a W-cycle. This multigrid process is effective if all smooth errors $\bfe$ satisfying \eqref{eq:algSmooth} are represented in the range of the prolongation $P$. For more information see \cite{BHM00,TOS01} and references therein.

\begin{algorithm}
\DontPrintSemicolon
\KwSty{Algorithm: $\bfu\leftarrow TwoGrid(H,\bfq,\bfu).$}\;
\begin{enumerate}\Indm
\item Apply pre-relaxations: $\bfu \leftarrow Relax(H,\bfu,\bfq)$\;
\item Compute and restrict the residual $\bfr_c = P^T(\bfq - H\bfu)$.
\item Compute $\bfe_c$ as the solution of the coarse grid problem $H_c\bfe_c=\bfr_c$.
\item Apply coarse grid correction: $\bfu \leftarrow \bfu + P\bfe_c$.
\item Apply post-relaxations: $\bfu \leftarrow Relax(H,\bfu,\bfq)$.
\end{enumerate}
\caption{Two-grid cycle.}
\label{alg:TwoCycle}
\end{algorithm}

To solve Helmholtz problems such as \eqref{eq:acousitcHelm} efficiently, the process above requires some modification, like in the shifted Laplacian framework. To apply this framework, one introduces a shifted matrix
\begin{equation}\label{eq:shift}
H_s = H - \im\alpha\omega^2M_s,
\end{equation}
where $H$ is the matrix defined by some discretization of the Helmholtz operator, $M_s$ is some mass matrix, and $\alpha>0$ is a shifting parameter. Usually, $M_s$ is defined as a mass matrix that is used for modeling attenuation in Helmholtz systems. In the acoustic case $M_s = \mbox{diag}(\bfkappa^2)$, and in our elastic case $M_s = A_f(\bfrho)$. The advantage of the shifted system is that it can be efficiently solved by multigrid methods. Hence, in the shifted Laplacian framework, the shifted Helmholtz matrix \eqref{eq:shift} is used as a preconditioner for a Helmholtz linear system \eqref{eq:linsys} inside a suitable Krylov method like (flexible) GMRES \cite{saad1993flexible}. 
The preconditioning is obtained by approximately inverting the shifted matrix \eqref{eq:shift} using a multigrid cycle.

When modeling waves with an attenuated Helmholtz problem (with $H_s$ instead of $H$), the waves decay rapidly if $\alpha$ is large. As we add more attenuation (larger $\alpha$ in \eqref{eq:shift}), we can invert the shifted matrix more easily, but the performance of the shifted Laplacian preconditioner deteriorates. This is a tradeoff that methods try to balance, and the common compromise chosen in \cite{erlangga2006novel} \emph{for the acoustic equation} is to use $\alpha = 0.5$. \cite{calandra2013improved} and \cite{JointEikFWI17} suggest using less attenuation, but invest more effort in the multigrid cycles. In \cite{rizzuti2016multigrid}  the \emph{elastic} equation \eqref{eq:elasticHelmSystem} is solved, and the authors use a high shift parameter ($\alpha=1.5$) in \eqref{eq:shift}, which results in a much less efficient preconditioner regardless of the multigrid ability to invert the shifted matrix.

The prolongation $P$ is usually chosen to be a bilinear interpolation operator, or an operator-induced prolongation like AMG \cite{erlangga2006novel}. As relaxation, the damped Jacobi method \cite{JointEikFWI17} or the GMRES method \cite{elman2001multigrid,calandra2013improved,cools2014new}, are often chosen for the acoustic case. For the elastic case, line-relaxation was used in \cite{rizzuti2016multigrid}. We elaborate on our choices of relaxation, interpolation and other parameters when we present our method in Section \ref{sec:Method}.

\subsection{Domain decomposition methods}

Domain decomposition (DD) \cite{gander2019class, dolean2015introduction} is a family of iterative methods that are based on decomposing the domain into subdomains in the continuous level, with or without overlapping. At each iteration, a local problem is solved on each sub-domain and the local solutions are attached on the interfaces, or in the overlapping regions, to construct a global solution. Let $\partial \Omega$ be the boundary of the entire domain and let $\partial \Omega_i$ be the boundary of the sub-domain $\Omega_i$. For every iteration, the boundary conditions on $\partial \Omega_i \cup \partial \Omega$ are inherited from the given boundary conditions of the entire domain. The interface conditions (ICs), i.e., the boundary conditions on $\partial \Omega_i \setminus \partial \Omega$ (see~\cite{dolean2015introduction}), are determined on each iteration using the previous approximation of the solution. 
In the serial version, given by Schwartz in~\cite{schwarz1870ueber}, we use the solution on the $(i-1)$-st domain to determine the ICs on the $i$-th domain, and in the parallel version, the solution of each sub-domain is preformed separately 
and then a global solution is constructed by averaging the local solutions on the overlaps. Finally, this global solution is used to determine the ICs of the next step.

\begin{figure}
  \centering
  \includegraphics[width=0.2\textwidth]{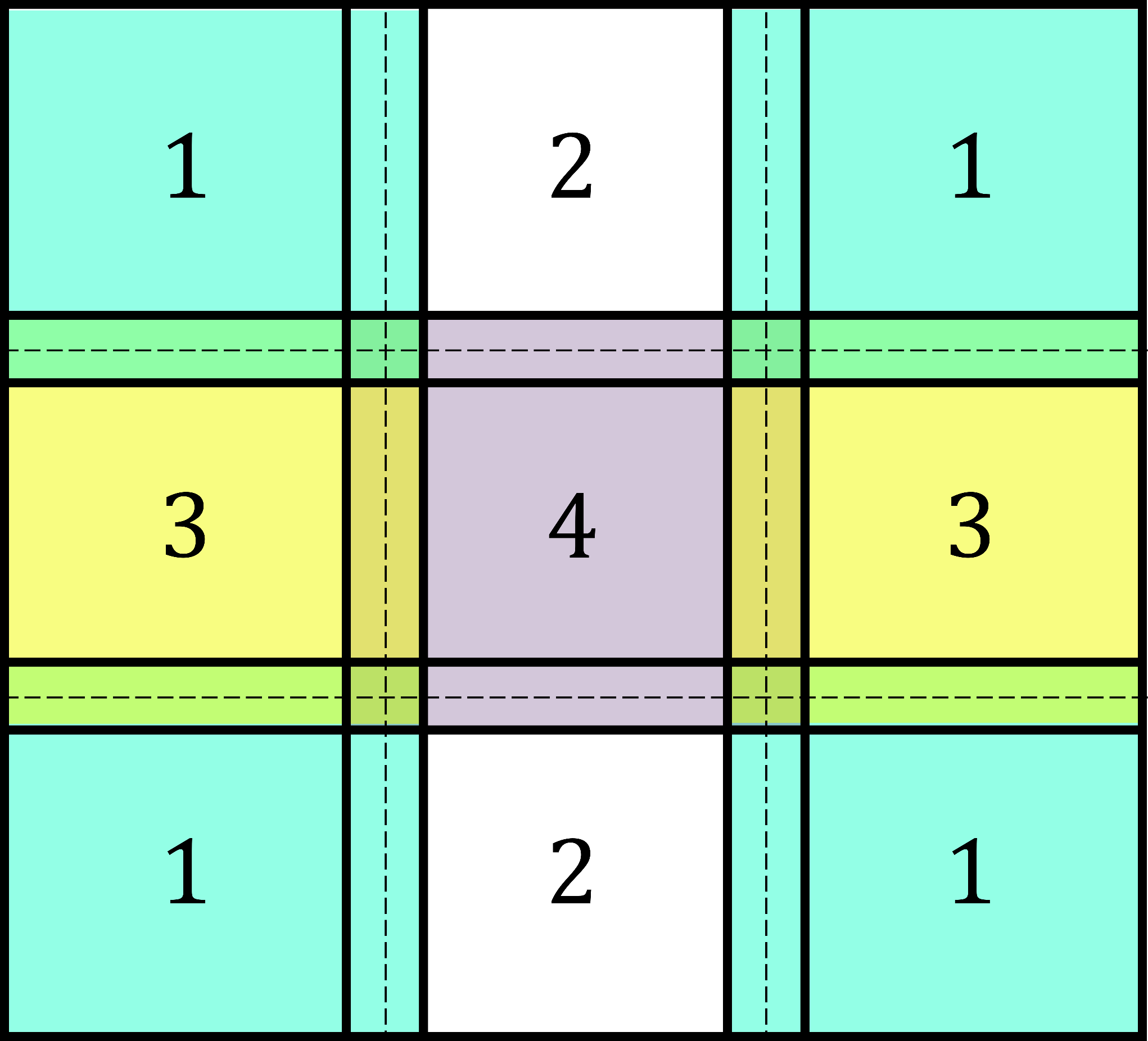}\\
  \caption{A coloring of the subdomains.}\label{fig:DDcolors}
\end{figure}

\textbf{Multi-coloring: } The advantage of the serial version is that the interface conditions on each step are more up to date. On the other hand, the serial version is less convenient to parallelize. 
We compensate between this advantage and the need for parallelism by using overlapping subdomains, and address them in a multicolor order, see Fig. \ref{fig:DDcolors}. That is, we partition the subdomains into several groups, denoted as colors. When updating domains of color number $1$, we only use the previous sweep to determine the ICs, but when updating domains of color number $4$, we already have full ICs given by the previous iterates during the same sweep. Algorithm \ref{alg:DD} summarizes the multicolored DD approach that we use.

To define the local differential operators, we use the approach of absorbing boundary conditions (ABCs) \cite{gander2013domain, graham2020domain} on the interfaces:
\begin{equation}
\frac{\partial \mathbf{u}}{\partial \mathbf{n}} - \im \beta \mathbf{u} = 0,
\end{equation}
only extend it to an absorbing boundary layer (ABLs) \cite{oosterlee2009shifted} like in \eqref{eq:elasticHelmMass}. The imaginary part of the ABLs grows gradually in the added layers outside of the physical domain. This is used to gradually damp the outgoing waves from within the subdomain.

\begin{algorithm}
\DontPrintSemicolon
\KwSty{Algorithm: $\vec{u}\leftarrow MulticolorDomainDecomposition(H,\bfq,\bfu).$}\;
\KwSty{Input: \normalfont{A collection of padded subdomains: }$\{\Omega_i\}_{i=1}^{n_{dom}}$}\;
\textit{\# $n_{colors}$: number of colors. The subdomains $\Omega_i$ of the same color are non-overlapping.}\;
\textit{\# $color(i)$: A function that returns the color of a subdomain $\Omega_i$.} \;
\textit{\# $H_{i}$: The discrete differential operator on the domain $\Omega_i$.}\;
\textit{\# $\bfr_i=\bfr_{\Omega_i}$: A vector $\bfr$ restricted to the subdomain $\Omega_i$.}\;
\For{$c=1,...,n_{colors}$}{
\textit{\# This inner loop can be done in parallel.}\;
\For{$i=1,...,n_{colors}\;\mbox{ such that }\;color(i)=c$}{
\begin{enumerate}\Indm
\item Compute the local residual $\bfr_i  = \bfq_i - (H\bfu)_{\Omega_i}$.
\item Solve for a local correction $H_i\bfe_i = \bfq_i$.
\item Update $\bfu_i \leftarrow \bfu_i + \bfe_i$.\vspace{-7pt}
\end{enumerate}
}
}
\caption{Multicolor domain decomposition method.}
\label{alg:DD}
\end{algorithm}

The order of solving the subdomains and the choice of the ICs has a huge influence on the convergence rate of DD methods. The recent \cite{taus2020sweeps} suggests an L-shaped order of sweeps. For non-overlapping DD, \cite{dai2022multidirectional, boubendir2012quasi} suggest different efficient solvers. Here we use the multicolor form of DD, which is quite basic, to complement our multigrid method. Other options for the domain sweeping are worthy of consideration as well.

\section{Shifted Laplacian multigrid for the elastic Helmholtz equation}\label{sec:Method}
It is clear that if we wish to solve a Helmholtz equation, whether \eqref{eq:acousitcHelm} or \eqref{eq:elasticHelm}, we first need to be able to solve the equation for a low frequency $\omega \approx 0$.
In \eqref{eq:acousitcHelm}, we are left with the elliptic weighted Poisson equation, which is considered to be the ``bread and butter'' of multigrid methods and can be easily solved even in cases of jumping coefficients and anisotropy. On the other hand, if we set $\omega = 0$ in \eqref{eq:elasticHelm} we get the linear elasticity equation which is also elliptic, but more difficult to solve than the Poisson equation. Thus, the idea that guides us is: if we wish to solve \eqref{eq:elasticHelm} or invert its shifted version, then our solver \emph{must} be able to handle the elasticity problem efficiently.

Linear elasticity problems can be solved by multigrid methods, but they require some special treatment in the nearly incompressible case when $\lambda \gg \mu$. In this case, the linear elasticity problem has a dominating grad-div operator, which has a rich null-space. Indeed, as $\div(\curl) = 0$, any vector function $\vec{v}$ which is a curl of another vector function $\vec{u}$ is in the null-space of the grad-div operator \cite{gaspar2008distributive}
\begin{equation}
\vec v = \curl \vec u \Rightarrow \nabla\div \vec v = 0.
\end{equation}
This equality also holds in the discrete space using staggered discretization. A simple prolongation operator cannot approximate this rich null-space well in its range, causing simple multigrid methods to be inefficient. Hence, a special treatment is required.

There are several approaches to handle this rich null-space. One is by using a prolongation that has this null-space in range, e.g. a prolongation based on smoothed aggregation which includes all the rigid body modes as basis functions \cite{SAV96}. This results in a rather large coarse grid matrix but does not require further modifications to the relaxation or other multigrid ingredients. A different family of approaches uses rather standard transfer operators, but also involves reformulating the system \eqref{eq:elasticHelmSystem} into an equivalent one, called ``mixed formulation'' \cite{gaspar2008distributive,zhu2010efficient}. The mixed formulation, which is the approach that we choose in this work, is achieved by introducing a new pressure variable $p = -(\lambda+\mu)\div\vec{u}$. In discrete form
\begin{equation}\label{eq:pressure}
\bfp = D_c\left(\bflambda+\bfmu\right)\gradh^T\vec\bfu,
\end{equation}
where the operator $D_c(\,)$ is the cell-centered diagonal matrix operator defined right after \eqref{eq:elasticHelmSystem}. We then reformulate the linear system \eqref{eq:elasticHelmSystem} as the coupled system
\begin{equation}\label{eq:elasticHelmReformulatedSystem}
\begin{pmatrix}
     \vec{\grad}_h^TA_e(\bfmu)\vec{\grad}_h - \omega^{2}M     & \gradh \\
  \gradh^T    & D_c(-\frac{1}{\bflambda+\bfmu})
\end{pmatrix}
\begin{pmatrix}
\vec\bfu\\
\bfp \end{pmatrix}
 =
\begin{pmatrix}
\vec\bfq\\
0 \end{pmatrix}.
\end{equation}
This linear system is equivalent to \eqref{eq:elasticHelmSystem}, because by construction, the system \eqref{eq:elasticHelmSystem} is the Schur complement of \eqref{eq:elasticHelmReformulatedSystem}, arises by eliminating the $\bfp$ block.

Using the reformulated system is the first step in tackling the problem. The second step is to use a special relaxation scheme. Here there are two options: either the so-called ``distributive relaxation'' \cite{gaspar2008distributive,zhu2010efficient}, or the cell-wise ``Vanka'' relaxation \cite{wobker2009numerical}. The latter was originally developed for the Stokes equation \cite{vanka1986block}, which has a similar (saddle-point) structure as \eqref{eq:elasticHelmReformulatedSystem} only with a zero block multiplying the $\bfp$ variable in the second equation. Both options are suitable for handling the problem, and in this work, we choose the cell-wise relaxation \cite{wobker2009numerical} which in our opinion is simpler to implement and to parallelize in multicore computations than the distributive relaxation.

Similarly to \cite{adler2016monolithic}, we can employ either full or economic versions of the cell-wise relaxation. In the full cell-wise relaxation, we sweep through all the cells in the domain, and for each cell we invert the local matrix composed of the block of \eqref{eq:elasticHelmReformulatedSystem} for that cell. In 3D, each cell has two variables for each displacement component $\bfu_i$, one on each of the faces, and the pressure variable at cell-center---a total of 7 variables, see Fig. \ref{fig:stag}. This leads to a $7\times7$ block of the matrix in \eqref{eq:elasticHelmReformulatedSystem} that is inverted for each cell. In the economical version of this method, instead of inverting the $7\times7$ block mentioned above, we take an easy-to-invert approximation where only the diagonal is considered for the $\bfu_i$ variables. We elaborate on this in Section \ref{sec:LFA}, in which we give local Fourier analysis (LFA) for the 2D case of this smoother. In 3D, this results in a storage of 19 variables per cell, and about the same floating operations per cell to apply the smoother.

\begin{figure}
  \centering
  \includegraphics[width=0.75\textwidth]{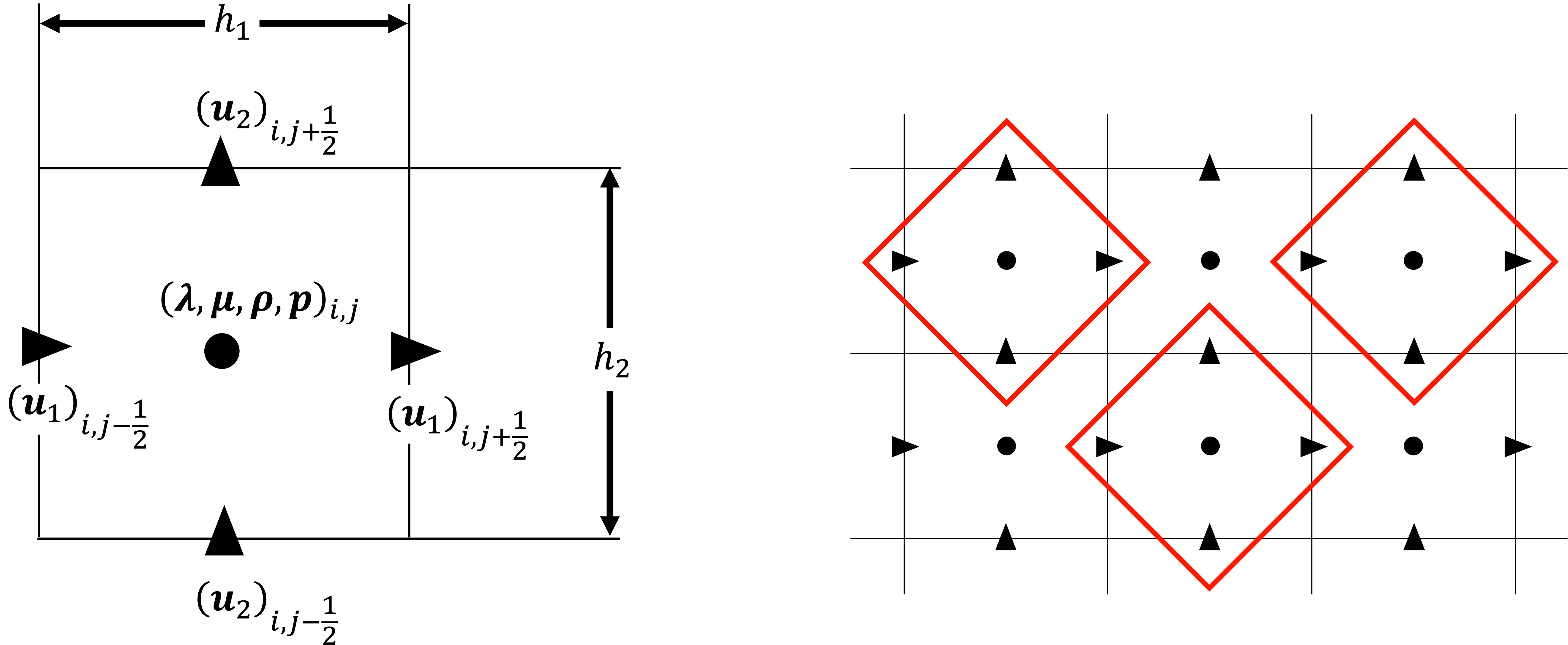}\\
  \caption{A two dimensional example of the variables that are relaxed together in a red-black cell-wise relaxation.}\label{fig:redBlack}
\end{figure}

Furthermore, the Vanka cell-wise smoothing can be applied by sweeping over the cells in a lexicographical order. This is analogous to the point-wise Gauss-Seidel, and hence Vanka called this method symmetric coupled Gauss-Seidel. In our LFA, in Section \ref{sec:LFA}, we refer to this ordering. However, this method is serial, and in practice we used red-black ordering to apply it in parallel.
In this ordering, all the cells in the domain are divided according to a checkerboard pattern into ``reds'' and ``blacks'', and we first simultaneously sweep over all the red cells, and then simultaneously sweep over all the black cells. Fig. \ref{fig:redBlack} shows an example of the cell-wise relaxation in red-black ordering in two dimensions. The lexicographic ordering is illustrated in our LFA in Section \ref{sec:LFA}---see Fig. \ref{FigBeforeAfter}.

\paragraph{Summary of the multigrid method} To solve the elastic Helmholtz equation we first reformulate the system \eqref{eq:elasticHelmSystem} to an equivalent mixed formulation system \eqref{eq:elasticHelmReformulatedSystem}. In our multigrid cycles we use standard bilinear transfer operators that are suitable for face-based staggered discretization \cite{TOS01}, and the operators on the coarser grid are defined by Galerkin coarsening. As relaxation we use damped red-black cell-wise relaxation. The multigrid hierarchy is defined for a shifted version of \eqref{eq:elasticHelmReformulatedSystem}, which is added with the zero-padded shift matrix
$$
\hat M_s = \begin{pmatrix}
     A_f(\bfrho)    & 0 \\
  0    & 0
\end{pmatrix}.
$$
That is, the artificial attenuation that we add involves only the $\vec\bfu$ block, just as we would have done for the original formulation \eqref{eq:elasticHelmSystem}. The attenuation is not involved with the pressure variable resulting from the mixed formulation. Finally, we solve the system \eqref{eq:elasticHelmReformulatedSystem} using a Krylov method preconditioned by a multigrid cycle for the shifted operator.

\begin{figure}
\begin{center}
	\newcommand{\image}[1]{\includegraphics[width=0.32\linewidth]{./images/#1}}
    \subfigure[\footnotesize Horizontal wavefield component $\bfu_1$.]{\image{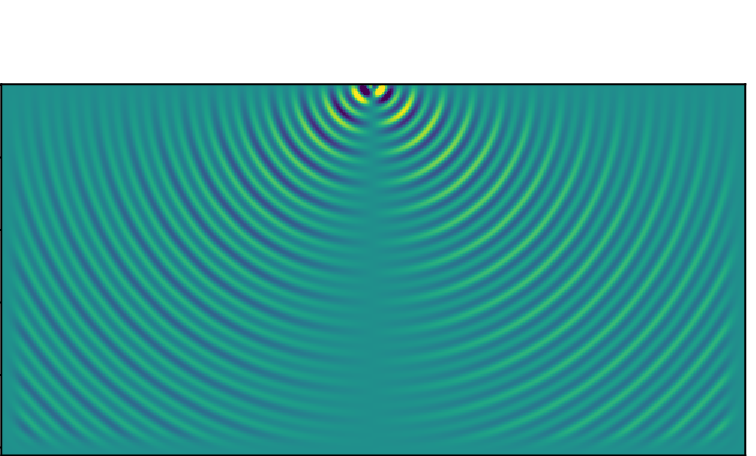}\label{fig:ux}}
    \subfigure[\footnotesize Vertical wavefield component $\bfu_2$]{\image{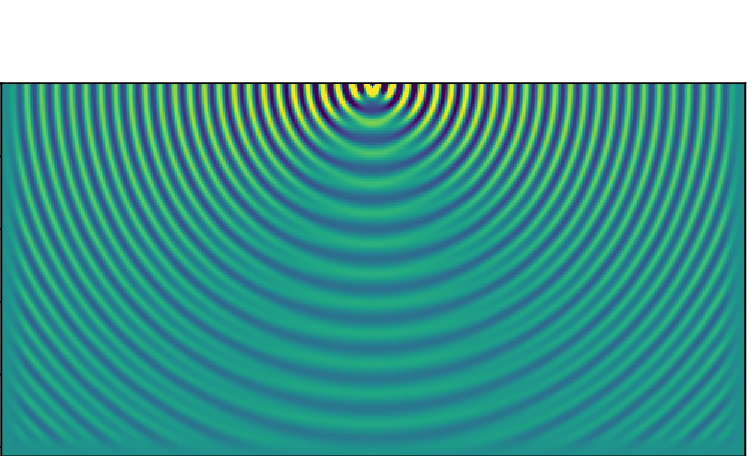}\label{fig:uz}}
    \subfigure[\footnotesize The pressure wavefield, $\bfp = -\gradh^\top \vec{\bfu}$.]{\image{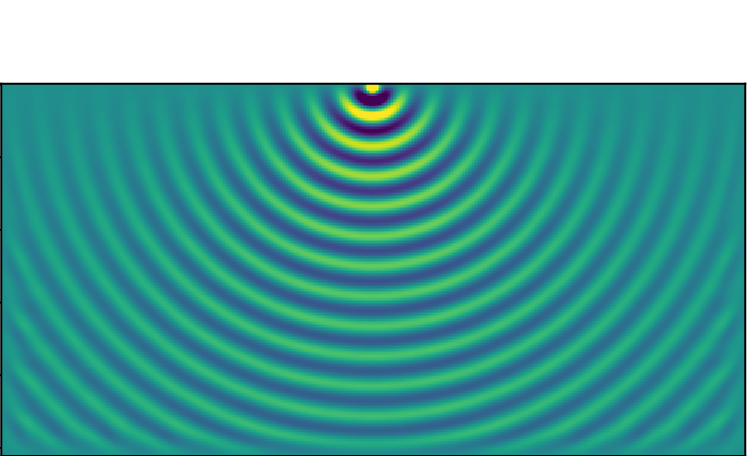}\label{fig:p}}
\end{center}
\caption{\footnotesize A two-dimentional wavefield for a point source, using an elastic medium with parameters $\lambda=2$, $\mu=\rho = 1$. 
}
\label{fig:wavefields}
\end{figure}

\paragraph{The relation between the acoustic and elastic Helmholtz equations} As noted before, if $\mu=0$, then the elastic equation models only pressure waves. Interestingly, the formulation \eqref{eq:elasticHelmReformulatedSystem} encapsulates this. The pressure variable introduced in \eqref{eq:pressure} is in fact the same pressure waveform function that appears in the acoustic \eqref{eq:acousitcHelm}, up to a diagonal scaling, see Fig. \ref{fig:wavefields} for an example of a wavefield for a point source. Indeed, if we set $\mu=0$ and ignore the attenuation parameter $\bfgamma$ in \eqref{eq:elasticHelmMass}, then the Schur complement of \eqref{eq:elasticHelmReformulatedSystem} when eliminating the $\vec{\bfu}$ block ends up as a cell-centered discretization of \eqref{eq:acousitcHelm}. This means that if we have a problem that is part acoustic and part elastic we can formulate the problem using one discrete system \eqref{eq:elasticHelmReformulatedSystem}, but treat both parts separately using the definition of $\bfp$ in \eqref{eq:pressure} and the elimination of $\vec\bfu$ from the top block of \eqref{eq:elasticHelmReformulatedSystem}, namely,
\begin{equation}
\vec\bfu = -\frac{1}{\omega^2}M^{-1}(\vec\bfq - \grad_h\vec\bfp).
\end{equation}
This results in a solver that reminds DD, in the sense that each domain is solved separately: the acoustic part is solved using standard shifted Laplacian, and the elastic part with elastic shifted Laplacian. The acoustic solver is obviously cheaper. Such scenarios of mixed elastic and acoustic media are common in marine full waveform inversion, where the sea is and acoustic medium and the rock under it is an elastic medium.

\paragraph{A remark on implementation and computational costs:} The cell-wise relaxation involves sweeping through all cells, and inverting a $7\times7$ matrix for each one. This may be a costly operation. Instead, in our implementation we extract these local matrices and invert them in the setup phase. Then, in the solve phase, which includes quite a few applications of the cell-wise relaxation, we only multiply the values of the inverted matrices (49 values per cell for 3D) instead of extracting and inverting the local matrices on-the-fly. In terms of storage (in 3D), this results in $49\times n_{cells}$ variables for every operator in the hierarchy. Note that the number of cells is approximately one fourth of the matrix variables for $\bfu_1,\bfu_2,\bfu_3,\bfp$. To reduce the storage costs, we convert and save the inverted matrices in a low 16-bit half precision. This results in quite fast application of the cell-wise relaxation in the price of a moderate storage requirement. Other ingredients like the coarsest grid factorization, and the vectors needed for the Krylov method are more memory consuming in our experience.

In terms of solve time, although box-smoothing requires more floating point operations (FLOPs) than point-wise smoothing, the difference is not dramatic. Recall that for any relaxation method, computing the residual of the elastic equations is needed. In 3D it requires at least 7 FLOPs per each variable. 
For point-wise relaxations, the application of a 7-point stencil for the Laplacian operator on each component $\bfu_i$ sums to a total of at least $21\times n_{cells}$ operations (depending on the implementation). 
Adding the multiplication of the div and grad operators in either \eqref{eq:elasticHelm} or \eqref{eq:elasticHelmReformulatedSystem}, the residual requires \emph{at least} four more FLOPs for each displacement component $\bfu_i,$ yielding a total of $33\times n_{cells}$. This is only 33\% less than the $49\times n_{cells}$ FLOPs that Vanka relaxation requires for multiplying a dense $7\times7$ matrix. Furthermore, in the economic version of the Vanka smoother, the matrix we invert has a favorable sparsity pattern with only 19 non-zeros and costs the same number of FLOPs to invert in our implementation. So, the economic version is cheaper in FLOPs than the standard residual calculation. Moreover, in both versions of the smoother the inversion of the matrix requires a continuous memory access, whereas in the residual computation, the stencil application requires non-continuous memory access patterns that are very costly in time.

\section{Hybrid domain decomposition and shifted Laplacian multigrid } \label{sec:hybridDD}

While the multigrid method presented in the previous section is the main component for our solution of the problem, it has two flaws. First, multigrid methods are not so easy to parallelize: coarser grids have less unknowns to distribute among workers, and the amount of inter-process communication is relatively high as it is required at the level of a matrix-vector product. The second drawback concerns the number of levels used in the multigrid hierarchy. The algebraically smooth error modes of the Helmholtz operator are not represented well on a very coarse grid, unlike other scenarios, because of their sign-changing patterns. As a result, the performance of the solver deteriorates as we use more levels at high frequency. The common choice is using three levels only \cite{calandra2013improved, JointEikFWI17, erlangga2008multilevel}, but invest more work on solving the second grid (e.g., by applying a W-cycle). However, using only three levels is problematic in large 3D cases, as we get a rather large coarsest grid problem, which is difficult to solve by a direct solver. Factorizing these matrices is highly memory consuming in 3D. These limitations and difficulties are amplified for the elastic system of equations compared to the acoustic scalar equation. The work in \cite{calandra2013improved} suggests an inexact solution of the coarsest grid using GMRES (for the acoustic case). This solution is very sensitive: it is rather expensive if one applies too many iterations, or does not perform well if one applies too few.

We address the two mentioned limitations, the need for parallelism and an efficient coarse solver, by harnessing the DD method to work together with our MG method. The idea is to exploit the inevitable local nature of both methods, to enjoy the advantages of each, with no significant degradation in the convergence properties of the preconditioner.

As a first step, we solve the coarse grid problem using DD. We call this method MG-DD. The local nature of the attenuated coarse system enables the division into subdomains, with no deterioration in the convergence of the multigrid method. Our second step, for improved parallelism, is using top-level DD and solve each subdomain by the suggested MG method. We call this method DD-MG. Using a multi-colored scheduling of the subdomain solutions as presented in Fig. \ref{fig:DDcolors}, we are able to trivially parallelize the subdomain MG solutions across several workers. Note that the system for each subdomain in the top level is not necessarily artificially attenuated. We add absorbing interface conditions (that mimic attenuation near the interfaces) as part of the DD method. However, to solve each top-level subdomain with MG, we add artificial attenuation as part of the MG preconditioner. Finally, we combine the two methods to a method named DD-MG-DD: we use DD both as a top-level preconditioner, and as a coarse grid solver in the resulting MG cycle within each subdomain. In our experiments, we approximately solve each subdomain by one multigrid W-cycle, but multiple cycles can be considered as well.

It is worth mentioning that more sophisticated DD preconditioners evolved in recent literature, compared to our multi-colored order DD, based on different sweeping order \cite{taus2020sweeps,dai2022multidirectional,leng2022trace} as well as special interface conditions \cite{boubendir2012quasi,stolk2013rapidly}. Some of which are less convenient with parallelism, and the comparison is beyond the scope of this paper. We demonstrate how a simple DD method can advance our MG method and leave further improvements for future research.
Although the combination with DD is essential for solving the elastic equation in 3D (using reasonable hardware), we note that the main contribution of this paper is the multigrid approach. Therefore, before presenting our numerical results for MG and DD combinations, in the next section we present the theoretical local Fourier analysis that justifies our MG algorithm that is used for each subdomain.

\section{Local Fourier analysis of the shifted Laplacian multigrid for the elastic Helmholtz equation} \label{sec:LFA}
In this section we provide a theoretical local Fourier analysis for the MG method presented in Section~\ref{sec:Method}, in 2D. We estimate the smoothing properties of the Vanka smoother applied on the elastic Helmholtz equation. We assume for the sake of the analysis that the Lam{\'e} coefficients $\mu(\vec{x}) = \mu$ and $\lambda(\vec{x}) = \lambda$ are constant, as well as the density $\rho(\vec{x}) = \rho.$ Thus, (\ref{eq:elasticHelmReformulatedSystem}) can be rewritten as:
\begin{equation} \label{eq:ElasticHelmholtzForLFA}
    H_h \begin{pmatrix}
        \mathbf{u} \\
        \mathbf{p}
    \end{pmatrix} =
    \begin{pmatrix}
        -\mu \vec{\Delta}_h - \omega^2 M & \nabla_h                \\
        \nabla^T_h                       & -\frac{1}{\lambda+\mu}I
    \end{pmatrix}
    \begin{pmatrix}
        \mathbf{u} \\
        \mathbf{p}
    \end{pmatrix}
    =
    \begin{pmatrix}
        \mathbf{q} \\
        \mathbf{0}
    \end{pmatrix}
\end{equation}
where the mass matrix is\footnote{Since the analysis in this section ignores the boundary conditions that are usually encapsulated in $\gamma$, we do not make a distinction between the physical and the added attenuation. We analyze an attenuated system, and denote the total attenuation by $\gamma$.}
\begin{equation}
    M = \rho (1-\im \gamma)I.
\end{equation}
In two dimensions, assuming staggered grid discretization, it reads
\begin{equation}
    H_h =
    \begin{pmatrix} \label{eq:DiscretizedHelmholtz}
        -\mu \Delta_h - \omega^2 M &                            & (\partial_u)_{h/2}      \\
                                   & -\mu \Delta_h - \omega^2 M & (\partial_v)_{h/2}      \\
        - (\partial_u)_{h/2}       & - (\partial_v)_{h/2}       & -\frac{1}{\lambda+\mu}I
    \end{pmatrix}
\end{equation}
where the shifted and weighted Laplacian is given by the stencil
\begin{equation} \label{eq:LStencil}
    - \mu \Delta_h - \omega^2 M =
    \begin{bmatrix}
                 & s_{0,1}  &         \\
        s_{-1,0} & s_{0,0}  & s_{1,0} \\
                 & s_{0,-1} &
    \end{bmatrix}
\end{equation}
with
\begin{equation}
    s_{0,0} = \frac{4\mu}{h^2} - \rho \omega^2 (1-\im\gamma)
    \quad \text{and} \quad s_{0,1} = s_{0,-1} = s_{1,0} = s_{-1,0} = - \frac{\mu}{h^2}
    ,\end{equation}
and the the central difference first derivatives are given by the stencils
\begin{equation} \label{eq:GStencil}
    (\partial_x)_{h/2} = \begin{bmatrix}
        s_{-\frac{1}{2},0} &
        0                  &
        s_{\frac{1}{2},0}
    \end{bmatrix}
    \quad
    \text{and}
    \quad
    (\partial_y)_{h/2} = \begin{bmatrix}
        s_{0,\frac{1}{2}} \\
        0                 \\
        s_{0,-\frac{1}{2}}
    \end{bmatrix}
\end{equation}
where
\begin{equation}
    s_{\frac{1}{2},0} = s_{0,\frac{1}{2}} = \frac{1}{h} \quad \text{and} \quad s_{-\frac{1}{2},0} = s_{0,-\frac{1}{2}} = -\frac{1}{h}
    .\end{equation}

\begin{figure}
    \centering
    \includegraphics[scale=0.15]{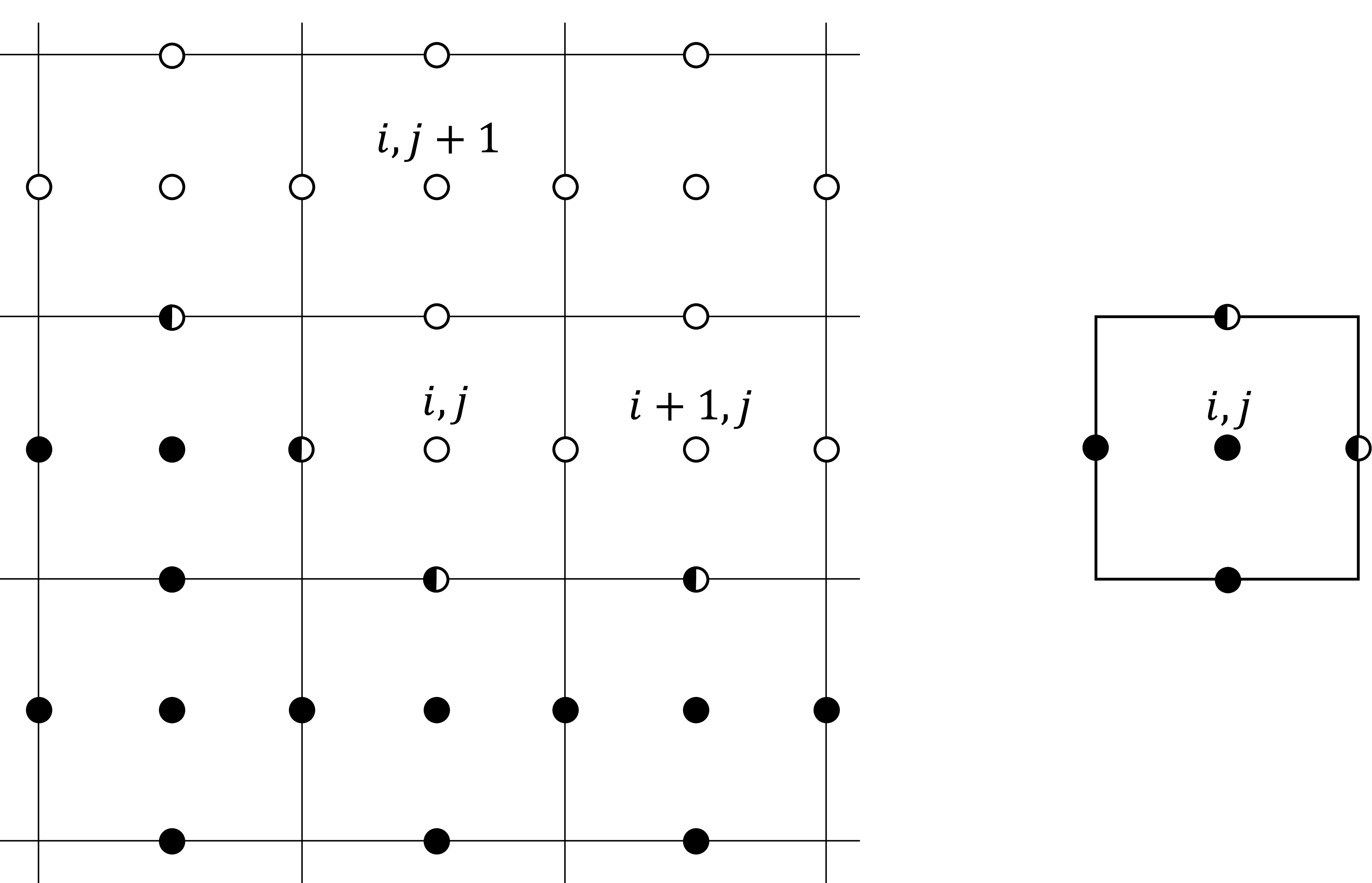}
    \caption{Update status before (on the left) and after (on the right) the relaxation on the $(i,j)$-th cell. $\Circle$ denotes an un-updated error, $\LEFTcircle$ denotes a partially updated error and $\CIRCLE$ denotes a fully updated error.} \label{FigBeforeAfter}
\end{figure}

We analyze the cell-wise relaxation that sweeps over the cells in a lexicographic order, see Fig.~\ref{FigBeforeAfter} for the update status of the grid right before and after the relaxation sweep over the $(i,j)$-th cell. 

\subsection{LFA preliminaries}
LFA is a predictive tool for the convergence of multigrid cycles. It was introduced by Brandt in~\cite{brandt1977multi}. This analysis works under two assumptions: a perfect coarse grid correction (a projection on the high frequencies), and a Toeplitz error propagation matrix (and thus diagonalizable by a Fourier basis). Under these two assumptions, the worst-case amplification of any frequency by the two-grid operator is nearly determined by the worst-case amplification of high frequencies only by the smoother. This gives rise to the definition of a smoothing factor for a system of equations (see, e.g.~\cite{trottenberg2000multigrid}, Chapter 8), given below.
\begin{definition} \label{def_mu_loc}
    Let $S_h$ be the error propagation matrix, and let $\tilde{S}_h(\theta)$ be its matrix of symbols, where $\theta\in[-\pi/2,3\pi/2]^2$. Then
    \begin{equation} \label{eq:muloc}
            \mu_{loc} \coloneqq \sup_{\theta\in T^{high}} \rho(\tilde{S}_h(\theta))
    \end{equation}
    where $T^{high}=[-\pi/2,3\pi/2]^2 \setminus [-\pi/2,\pi/2]^2$.
\end{definition}

For a point-wise relaxation, $S_h$ is calculated by a simple splitting of the original operator. However, for cell-wise overlapping smoothers, $S_h$ can not be easily calculated by a splitting. In~\cite{sivaloganathan1991use}, a local mode analysis for a lexicographic order Vanka smoother is given for the Stokes equations in two dimensions. We analyze the damped elastic Helmholtz equation similarly. A two-grid analysis for the Stokes equations with finite elements discretization is given in \cite{farrell2021local}. There, the authors used additive Vanka (rather than multiplicative), and offered an approach of different damping parameter for each component. In the case of overlapping smoothers, it is not immediately clear why $S_h$ does not intermix Fourier modes. In~\cite{maclachlan2011local}, it is proved for a class of overlapping smoothers, giving our analysis a theoretical justification. 

When the coarse grid correction is nearly ideal, the smoothing factor gives a good prediction for the convergence of the multigrid cycle. However, when this is not the case, the two-grid factor can give a more valuable prediction. 

Let $TG = S^{\nu_2} (I - PA_c^{-1}RA_f) S^{\nu_1}$ be the two-grid operator (described in Algorithm \ref{alg:TwoCycle}) when $A_f$ is the original operator discretized on a fine grid, $A_c$ is its coarse grid approximation (in our case, calculated by the Galerkin product $A_c = R A_f P$), $R$ and $P$ are the restriction and prolongation and $\nu_1,\nu_2$ is the number of pre- and post-smoothing. The definition of the two-grid factor (see \cite{trottenberg2000multigrid}) is given bellow:
\begin{definition} \label{def_rho_loc}
     The two-grid factor is defined as:
    \[
        \rho_{loc} \coloneqq \sup_{\theta\in T^{low}} \rho(\widetilde{TG}(\theta))
    \]
   where $\widetilde{TG}(\theta)$ is the matrix of symbols of the two-grid operator and
   $T^{low}=\left[\frac{\pi}{2},\frac{\pi}{2}\right]^2$. 
\end{definition}
We omit the details about the calculation of the symbol matrix $\widetilde{TG}$. For a more detailed calculation of the two-grid operator for the (scalar) acoustic Helmholtz equation, see \cite{cools2013local}, and for a two-grid analysis of the Stokes system of equations see \cite{farrell2021local}. We apply a similar analysis here, in combination with our smoothing analysis for the multiplicative Vanka smoother, that we give in detail bellow.

\subsection{Notation and method for the smoothing analysis}
Let $A$ be the $5\times 5$ sub-matrix of the discretized operator $H_h$ from \eqref{eq:DiscretizedHelmholtz}, containing only the DOF's of the $(i,j)$-th cell. Let $B$ be an easy-to-invert approximation of $A$. Let~$\mathbf{e}^{(k)}$ be the $5\times 1$ vector of errors after the $k$-th relaxation step, and let $\mathbf{r}^{(k)}$ be the corresponding vector of residuals. Assuming that we sweep over the $(i,j)$-th cell in the $(k+1)$-st relaxation step, we update $\mathbf{e}^{(k)}$ as following:
\begin{equation} \label{eq:update_error}
    \mathbf{e}^{(k+1)} = \mathbf{e}^{(k)} - w B^{-1} \mathbf{r}^{(k)}
\end{equation}
where $w$ is a damping parameter. In terms of corrections, the residuals are therefore
\begin{equation} \label{eq:relaxation}
    \mathbf{r}^{(k)} \coloneqq A\mathbf{e}^{(k)} = -B(\mathbf{e}^{(k+1)} - \mathbf{e}^{(k)})/w.
\end{equation}

For the ease of notation, let $u,v,p$ denote the unknowns we previously denoted by $u_1,u_2,p$. In order to write the system of equations~\eqref{eq:relaxation} explicitly, we denote the vectors of errors, corrections and residuals by
\begin{align}
    \mathbf{e}^{(k)} & =
    (
    e^u_{i-\frac{1}{2},j},
    e^u_{i+\frac{1}{2},j},
    e^v_{i,j-\frac{1}{2}},
    e^v_{i,j+\frac{1}{2}},
    e^p_{i,j}
    )^T,
    \\
    \mathbf{c}^{(k)} & = \mathbf{e}^{(k+1)} - \mathbf{e}^{(k)} =
    (
    c^u_{i-\frac{1}{2},j},
    c^u_{i+\frac{1}{2},j},
    c^v_{i,j-\frac{1}{2}},
    c^v_{i,j+\frac{1}{2}},
    c^p_{i,j}
    )^T
    \\
    \mathbf{r}^{(k)} & =
    (
    r^u_{i-\frac{1}{2},j},
    r^u_{i+\frac{1}{2},j},
    r^v_{i,j-\frac{1}{2}},
    r^v_{i,j+\frac{1}{2}},
    r^p_{i,j}
    )^T
\end{align}
respectively. With this notation, Equation~\eqref{eq:relaxation} can be rewritten as
\begin{equation} \label{eq:RelaxationMatrix}
    - \underbrace{\begin{pmatrix}
            s_{0,0}             &                    &                     &                    & s_{\frac{1}{2},0}        \\
                                & s_{0,0}            &                     &                    & s_{-\frac{1}{2},0}       \\
                                &                    & s_{0,0}             &                    & s_{0,\frac{1}{2}}        \\
                                &                    &                     & s_{0,0}            & s_{0,-\frac{1}{2}}       \\
            -s_{-\frac{1}{2},0} & -s_{\frac{1}{2},0} & -s_{0,-\frac{1}{2}} & -s_{0,\frac{1}{2}} & -\frac{1}{\lambda + \mu}
        \end{pmatrix}}_B
    \begin{pmatrix}
        c^u_{i-\frac{1}{2},j}/w_u \\
        c^u_{i+\frac{1}{2},j}/w_u \\
        c^v_{i,j-\frac{1}{2}}/w_v\\
        c^v_{i,j+\frac{1}{2}}/w_v \\
        c^p_{i,j}/w_p
    \end{pmatrix}
    = \begin{pmatrix}
        r^u_{i-\frac{1}{2},j} \\
        r^u_{i+\frac{1}{2},j} \\
        r^v_{i,j-\frac{1}{2}} \\
        r^v_{i,j+\frac{1}{2}} \\
        r^p_{i,j}
    \end{pmatrix}
\end{equation}
where $w_u,w_v,w_p$ are the damping parameters for each component (as we show bellow, choosing the same damping for each equation in this system is not necessarily optimal).

Let us give a notation for the error in the frequency domain, assuming that the error is comprised of a single Fourier mode. Before the relaxation sweep, the error will be denoted as
\begin{equation} \label{eq:unupdate}
    \begin{pmatrix}
        e_u^{\Tiny{\Circle}} \\
        e_v^{\Tiny{\Circle}}  \\
        e_p^{\Tiny{\Circle}}
    \end{pmatrix}
    = \begin{pmatrix}
        \alpha_u^{\Tiny{\Circle}}(\theta) \\
        \alpha_v^{\Tiny{\Circle}}(\theta) \\
        \alpha_p^{\Tiny{\Circle}}(\theta)
    \end{pmatrix}
    e^{\textit{\i} \theta \cdot \mathbf{x} /h}
\end{equation}
where $\mathbf{x} = [ih,jh]^T$ is the location of the center of the $(i,j)$-th cell. The displacement components' errors are corrected twice, since each face belongs to two cells, see Fig.~\ref{FigBeforeAfter}. The pressure's error is corrected once. We denote the partially and fully corrected errors by
\begin{equation} 
    \begin{pmatrix}
        e_u^{\Tiny{\LEFTcircle}} \\
        e_v^{\Tiny{\LEFTcircle}}
    \end{pmatrix}
    =\begin{pmatrix}
        \alpha_u^{\Tiny{\LEFTcircle}}(\theta) \\
        \alpha_v^{\Tiny{\LEFTcircle}}(\theta)
    \end{pmatrix}
    e^{\textit{\i} \theta \cdot \mathbf{x} /h}
 ,\quad \mbox{and}\quad 
    \begin{pmatrix}
        e_u^{\Tiny{\CIRCLE}} \\
        e_v^{\Tiny{\CIRCLE}} \\
        e_p^{\Tiny{\CIRCLE}}
    \end{pmatrix}
    = \begin{pmatrix}
        \alpha_u^{\Tiny{\CIRCLE}}(\theta) \\
        \alpha_v^{\Tiny{\CIRCLE}}(\theta) \\
        \alpha_p^{\Tiny{\CIRCLE}}(\theta)
    \end{pmatrix}
    e^{\textit{\i} \theta \cdot \mathbf{x} /h},
\end{equation}
respectively. Since we assume a single mode, we can omit the dependence in $\theta$ and denote, e.g., $\alpha_u^{\Tiny{\Circle}} = \alpha_u^{\Tiny{\Circle}}(\theta)$. Our goal is to find a matrix $\tilde{S}_h=\tilde{S}_h(\theta)$ such that
\begin{equation} \label{eq:Sh_wishful}
    \begin{pmatrix}
        \alpha_u^{\Tiny{\CIRCLE}} \\
        \alpha_v^{\Tiny{\CIRCLE}} \\
        \alpha_p^{\Tiny{\CIRCLE}}
    \end{pmatrix}
    = \tilde{S}_h
    \begin{pmatrix}
        \alpha_u^{\Tiny{\Circle}} \\
        \alpha_v^{\Tiny{\Circle}} \\
        \alpha_p^{\Tiny{\Circle}}
    \end{pmatrix}.
\end{equation}
However, the fully corrected errors are influenced not only by the initial errors, but also by the partially corrected ones. Thus, following~\cite{sivaloganathan1991use}, we first find a~$5\times 5$ matrix $P=P(\theta)$ and a~$5\times 3$ matrix $Q=Q(\theta)$ such that
\begin{equation} \label{eq:PQdef}
    P
    \begin{pmatrix}
        \alpha_u^{\Tiny{\LEFTcircle}} \\
        \alpha_v^{\Tiny{\LEFTcircle}} \\
        \alpha_u^{\Tiny{\CIRCLE}}     \\
        \alpha_v^{\Tiny{\CIRCLE}}     \\
        \alpha_p^{\Tiny{\CIRCLE}}
    \end{pmatrix}
    = Q
    \begin{pmatrix}
        \alpha_u^{\Tiny{\Circle}} \\
        \alpha_v^{\Tiny{\Circle}} \\
        \alpha_p^{\Tiny{\Circle}}
    \end{pmatrix}
\end{equation}
and then define $\tilde{S}_h$ to be the lower~$3\times 3$ block of $P^{-1}Q,$ namely
\begin{equation} \label{eq:Sh_def}
    \begin{pmatrix}
        \alpha_u^{\Tiny{\LEFTcircle}} \\
        \alpha_v^{\Tiny{\LEFTcircle}} \\
        \alpha_u^{\Tiny{\CIRCLE}}     \\
        \alpha_v^{\Tiny{\CIRCLE}}     \\
        \alpha_p^{\Tiny{\CIRCLE}}
    \end{pmatrix}
    = \begin{pmatrix}
        \star \\
        \tilde{S}_h
    \end{pmatrix}
    \begin{pmatrix}
        \alpha_u^{\Tiny{\Circle}} \\
        \alpha_v^{\Tiny{\Circle}} \\
        \alpha_p^{\Tiny{\Circle}}
    \end{pmatrix}.
\end{equation}
The spectral radius of $\tilde{S}_h(\theta)$ gives the amplification factor as a function of $\theta$, and the worst amplification factor over $\theta\in T^{high}$ is the smoothing factor $\mu_{loc}$, see Definition~\ref{def_mu_loc}.

In the case of non-overlapping smoothers, the symbol can be calculated for a general location on the grid. Here, the overlaps between cells demands a location-dependant notation. Recall that \eqref{eq:LStencil} applied on the Fourier mode $e^{\im \theta \cdot \mathbf{x} /h}$ yields
\begin{equation} \label{eq:symbol_laplace}
    (s_{0,0} + s_{0,1} e^{\im\theta_2} + s_{0,-1} e^{-\im\theta_2} + s_{1,0} e^{\im\theta_1} + s_{-1,0} e^{-\im\theta_1}) e^{\im \theta \cdot \mathbf{x} /h}.
\end{equation}
A similar observation is true for the stencil~(\ref{eq:GStencil}). In this spirit, we denote
\begin{equation} \label{eq:tildeS}
    \tilde{s}_{l,m} \coloneqq s_{l,m} e^{\textit{\i} (l\theta_1 + m\theta_2)} e^{\textit{\i} \theta \cdot \mathbf{x} / h},
\end{equation}
where $l,m \in \{\pm 1, \pm 1/2, 0\}$, e.g., $\tilde{s}_{0,-\frac{1}{2}} = s_{0,-\frac{1}{2}} e^{-\textit{\i} \theta_2 / 2} e^{\textit{\i} \theta \cdot \mathbf{x} / h}$ and $\tilde{s}_{0,0} = s_{0,0} e^{\textit{\i} \theta \cdot \mathbf{x} / h}$.

\subsection{Calculation of the smoothing factor}
We calculate the smoothing factor $\mu_{loc}$ for the Vanka relaxation applied on \eqref{eq:DiscretizedHelmholtz}.

In terms of Fourier modes, using Fig.~\ref{FigBeforeAfter}, \eqref{eq:RelaxationMatrix} reads
\begin{align} \label{eq:i-1/2,j}
    r^u_{i-\frac{1}{2},j} & = -\textstyle{\frac{1}{w}} e^{-\im \theta_1 /2} \left(\tilde{s}_{0,0} (\alpha_u^{\Tiny{\CIRCLE}} - \alpha_u^{\Tiny{\LEFTcircle}}) + \tilde{s}_{\frac{1}{2},0} (\alpha_p^{\Tiny{\CIRCLE}} - \alpha_p^{\Tiny{\Circle}}) \right)
\end{align}
\begin{align} \label{eq:i+1/2,j}
    r^u_{i+\frac{1}{2},j} & = - \textstyle{\frac{1}{w}} e^{\im \theta_1 /2} \left(\tilde{s}_{0,0} (\alpha_u^{\Tiny{\LEFTcircle}} - \alpha_u^{\Tiny{\Circle}}) + \tilde{s}_{-\frac{1}{2},0} (\alpha_p^{\Tiny{\CIRCLE}} - \alpha_p^{\Tiny{\Circle}})\right)
\end{align}
\begin{align}
    \label{eq:i,j-1/2}
    r^v_{i,j-\frac{1}{2}} & = -\textstyle{\frac{1}{w}} e^{-\im \theta_2 /2} \left(\tilde{s}_{0,0} (\alpha_v^{\Tiny{\CIRCLE}} - \alpha_v^{\Tiny{\LEFTcircle}}) + \tilde{s}_{0,\frac{1}{2}} (\alpha_p^{\Tiny{\CIRCLE}} - \alpha_p^{\Tiny{\Circle}}) \right)
\end{align}
\begin{align} \label{eq:i,j+1/2}
   r^v_{i,j+\frac{1}{2}} & = - \textstyle{\frac{1}{w}} e^{\im \theta_2 /2} \left(\tilde{s}_{0,0} (\alpha_v^{\Tiny{\LEFTcircle}} - \alpha_v^{\Tiny{\Circle}}) + \tilde{s}_{0,-\frac{1}{2}} (\alpha_p^{\Tiny{\CIRCLE}} - \alpha_p^{\Tiny{\Circle}}) \right)
\end{align}
\begin{align} \label{eq:i,j}
    r^p_{i,j} & = \frac{1}{w} \Big(
    + \tilde{s}_{-\frac{1}{2},0} (\alpha_u^{\Tiny{\CIRCLE}} - \alpha_u^{\Tiny{\LEFTcircle}})
    + \tilde{s}_{\frac{1}{2},0} (\alpha_u^{\Tiny{\LEFTcircle}} - \alpha_u^{\Tiny{\Circle}})                                                                                                                                                                                                                 \\ \nonumber
              & \qquad \qquad \, \, + \tilde{s}_{0,-\frac{1}{2}} (\alpha_v^{\Tiny{\CIRCLE}} - \alpha_v^{\Tiny{\LEFTcircle}}) + \tilde{s}_{0,\frac{1}{2}} (\alpha_v^{\Tiny{\LEFTcircle}} + \alpha_v^{\Tiny{\Circle}}) + \frac{1}{\lambda+\mu} (\alpha_p^{\Tiny{\CIRCLE}} - \alpha_p^{\Tiny{\Circle}}) \Big).
\end{align}

Writing $\mathbf{r}^{(k)} = A\mathbf{e}^{(k)}$ using the stencils~(\ref{eq:LStencil}) and~(\ref{eq:GStencil}), we have
\begin{align}
    r^u_{i-\frac{1}{2},j} & = s_{0,0} e^u_{i-\frac{1}{2},j} + s_{0,1} e^u_{i-\frac{1}{2},j+1} + s_{0,-1} e^u_{i-\frac{1}{2},j-1} + s_{1,0} e^u_{i+\frac{1}{2},j} + s_{-1,0} e^u_{i-\frac{3}{2},j} +s_{\frac{1}{2},0} e^p_{i,j} + s_{-\frac{1}{2},0} e^p_{i-1,j}
    \\[0.6em]
    r^u_{i+\frac{1}{2},j} & =  s_{0,0} e^u_{i+\frac{1}{2},j} + s_{0,1} e^u_{i+\frac{1}{2},j+1} + s_{0,-1} e^u_{i+\frac{1}{2},j-1} +s_{1,0} e^u_{i+\frac{3}{2},j} + s_{-1,0} e^u_{i-\frac{1}{2},j} + s_{\frac{1}{2},0} e^p_{i+1,j} + s_{-\frac{1}{2},0} e^p_{i,j}
    \\[0.6em]
    r^v_{i,j-\frac{1}{2}} & = s_{0,0} e^v_{i,j-\frac{1}{2}} + s_{0,1} e^v_{i,j+\frac{1}{2}} + s_{0,-1} e^v_{i, j-\frac{3}{2}} + s_{1,0} e^v_{i+1,j-\frac{1}{2}} + s_{-1,0} e^v_{i-1,j-\frac{1}{2}} + s_{0,\frac{1}{2}} e^p_{i,j} + s_{0,-\frac{1}{2}} e^p_{i,j-1}
    \\[0.6em]
    r^v_{i,j+\frac{1}{2}} & = s_{0,0} e^v_{i,j+\frac{1}{2}} + s_{0,1} e^v_{i,j+\frac{3}{2}} + s_{0,-1} e^v_{i,j-\frac{1}{2}} + s_{1,0} e^v_{i+1,j+\frac{1}{2}} + s_{-1,0} e^v_{i-1,j+\frac{1}{2}} + s_{0, \frac{1}{2}} e^p_{i,j+1} + s_{0,-\frac{1}{2}} e^p_{i,j}
    \\
    r^p_{i,j}             & = - s_{\frac{1}{2},0} e^u_{i+\frac{1}{2},j} - s_{-\frac{1}{2},0} e^u_{i-\frac{1}{2},j} -  s_{0,\frac{1}{2}} e^v_{i,j+\frac{1}{2}} - s_{0,-\frac{1}{2}} e^v_{i,j-\frac{1}{2}} - \frac{1}{\lambda + \mu} e^p_{i,j}.
\end{align}
To write it in terms of Fourier modes, we use Fig.~\ref{FigBeforeAfter}. For instance, for the left face of the~$(i,j)$-th cell, located in $[(i-1/2)h,jh]^T$, we multiply each $\tilde{s}_{l,m}$ by $e^{-\textit{\i} \theta_1 /2}.$ Recall that the notation $\tilde{s}_{l,m}$ already includes the product by $e^{\textit{\i} \theta \cdot \mathbf{x} /h}$, which locates us in the center of the $(i,j)$-th cell. We therefore get
\begin{align} \label{eq:before_i-1/2,j}
    r^u_{i-\frac{1}{2},j} & = e^{-\textit{\i} \theta_1 /2} \left(\tilde{s}_{0,0} \alpha_u^{\Tiny{\LEFTcircle}} + \tilde{s}_{0,1} \alpha_u^{\Tiny{\Circle}} + \tilde{s}_{0,-1} \alpha_u^{\Tiny{\CIRCLE}} + \tilde{s}_{1,0} \alpha_u^{\Tiny{\Circle}} + \tilde{s}_{-1,0} \alpha_u^{\Tiny{\CIRCLE}} + \tilde{s}_{\frac{1}{2},0} \alpha_p^{\Tiny{\Circle}} + \tilde{s}_{-\frac{1}{2},0} \alpha_p^{\Tiny{\CIRCLE}}\right)       \\[0.6em] \label{eq:before_i+1/2,j}
    r^u_{i+\frac{1}{2},j} & = e^{\textit{\i} \theta_1 /2} \left( \tilde{s}_{0,0} \alpha_u^{\Tiny{\Circle}} + \tilde{s}_{0,1} \alpha_u^{\Tiny{\Circle}} + \tilde{s}_{0,-1} \alpha_u^{\Tiny{\CIRCLE}} + \tilde{s}_{1,0} \alpha_u^{\Tiny{\Circle}} + \tilde{s}_{-1,0} \alpha_u^{\Tiny{\LEFTcircle}} + \tilde{s}_{\frac{1}{2},0} \alpha_p^{\Tiny{\Circle}} + \tilde{s}_{-\frac{1}{2},0} \alpha_p^{\Tiny{\Circle}} \right)      \\[0.6em] \label{eq:before_i,j-1/2}
    r^v_{i,j-\frac{1}{2}} & = e^{-\textit{\i} \theta_2 /2} \left( \tilde{s}_{0,0} \alpha_v^{\Tiny{\LEFTcircle}} + \tilde{s}_{0,1} \alpha_v^{\Tiny{\Circle}} + \tilde{s}_{0,-1} \alpha_v^{\Tiny{\CIRCLE}} + \tilde{s}_{1,0} \alpha_v^{\Tiny{\LEFTcircle}} + \tilde{s}_{-1,0} \alpha_v^{\Tiny{\CIRCLE}} + \tilde{s}_{0,\frac{1}{2}} \alpha_p^{\Tiny{\Circle}} + \tilde{s}_{0,-\frac{1}{2}} \alpha_p^{\Tiny{\CIRCLE}} \right) \\[0.6em] \label{eq:before_i,j+1/2}
    r^v_{i,j+\frac{1}{2}} & = e^{\textit{\i} \theta_2 /2} \left( \tilde{s}_{0,0} \alpha_v^{\Tiny{\Circle}} + \tilde{s}_{0,1} \alpha_v^{\Tiny{\Circle}} + \tilde{s}_{0,-1} \alpha_v^{\Tiny{\LEFTcircle}} + \tilde{s}_{1,0} \alpha_v^{\Tiny{\Circle}} + \tilde{s}_{-1,0} \alpha_v^{\Tiny{\LEFTcircle}} + \tilde{s}_{0, \frac{1}{2}} \alpha_p^{\Tiny{\Circle}} + \tilde{s}_{0,-\frac{1}{2}} \alpha_p^{\Tiny{\Circle}} \right) \\ \label{eq:before_i,j}
    r^p_{i,j}             & =  - \tilde{s}_{\frac{1}{2},0} \alpha_u^{\Tiny{\Circle}} - \tilde{s}_{-\frac{1}{2},0} \alpha_u^{\Tiny{\LEFTcircle}} - \tilde{s}_{0,\frac{1}{2}} \alpha_v^{\Tiny{\Circle}} - \tilde{s}_{0,-\frac{1}{2}} \alpha_v^{\Tiny{\LEFTcircle}} - \frac{1}{\lambda + \mu} \alpha_p^{\Tiny{\Circle}} .
\end{align}

Now we can write a relation of the form~\eqref{eq:PQdef}. Equating \eqref{eq:before_i-1/2,j} and \eqref{eq:i-1/2,j} and rearranging gives
\begin{equation}\label{eq:PQcalc}
    p_{1,1} \alpha_u^{\Tiny{\LEFTcircle}} + p_{1,3} \alpha_u^{\Tiny{\CIRCLE}} + p_{1,5} \alpha_p^{\Tiny{\CIRCLE}}
    = q_{1,1} \alpha_u^{\Tiny{\Circle}} + q_{1,3} \alpha_p^{\Tiny{\Circle}}
\end{equation}
where
\begin{align}
    p_{1,1} & \coloneqq \left(1-1/w_u\right)\tilde{s}_{0,0}                            &
    p_{1,3} & \coloneqq \tilde{s}_{0,-1}+\tilde{s}_{-1,0} + (1/w_u) \tilde{s}_{0,0}    &
    p_{1,5} & \coloneqq \tilde{s}_{-\frac{1}{2},0} + (1/w_u) \tilde{s}_{\frac{1}{2},0}   \\
    q_{1,1} & \coloneqq - \tilde{s}_{0,1} - \tilde{s}_{1,0}                          &
    q_{1,3} & \coloneqq  \left(-1+1/w_u\right)\tilde{s}_{\frac{1}{2},0}.
\end{align}
Continuing the process, equating \eqref{eq:before_i+1/2,j} and \eqref{eq:i+1/2,j}, we get
\begin{align}
    p_{2,1} & \coloneqq \tilde{s}_{-1,0} + (1/w_u)\tilde{s}_{0,0}                                          &
    p_{2,3} & \coloneqq \tilde{s}_{0,-1}                                                                 &
    p_{2,5} & \coloneqq (1/w_u)\tilde{s}_{-\frac{1}{2},0}                                                           \\
    q_{2,1} & \coloneqq \left(-1+1/w_u\right)\tilde{s}_{0,0}-\tilde{s}_{0,1}-\tilde{s}_{1,0}         &
    q_{2,3} & \coloneqq -\tilde{s}_{\frac{1}{2},0} + \left(-1+1/w_u\right)\tilde{s}_{-\frac{1}{2},0},
\end{align}

equating \eqref{eq:before_i,j-1/2} and \eqref{eq:i,j-1/2} yields
 \begin{align}
    p_{3,2} & \coloneqq \left(1-1/w_v\right) \tilde{s}_{0,0} + \tilde{s}_{1,0}                             &
    p_{3,4} & \coloneqq \tilde{s}_{0,-1} + \tilde{s}_{-1,0} + (1/w_v) \tilde{s}_{0,0}  &                    
    p_{3,5} & \coloneqq \tilde{s}_{0,-\frac{1}{2}} + (1/w_v) \tilde{s}_{0,\frac{1}{2}}      \\                  
    q_{3,2} & \coloneqq -\tilde{s}_{0,1}                                                           &
    q_{3,3} & \coloneqq \left(-1+1/w_v\right)\tilde{s}_{0,\frac{1}{2}},
\end{align}
equating \eqref{eq:before_i,j+1/2} and \eqref{eq:i,j+1/2} results in
\begin{align}
    p_{4,2} & \coloneqq \tilde{s}_{0,-1} + \tilde{s}_{-1,0} + (1/w_v) \tilde{s}_{0,0}                      &
    p_{4,5} & \coloneqq (1/w_v)\tilde{s}_{0,-\frac{1}{2}}                                                           \\
    q_{4,2} & \coloneqq \left(-1+1/w_v\right)\tilde{s}_{0,0} - \tilde{s}_{0,1} - \tilde{s}_{1,0}     &
    q_{4,3} & \coloneqq -\tilde{s}_{0,\frac{1}{2}} - \left(1-1/w_v\right) \tilde{s}_{0,-\frac{1}{2}}
\end{align}
and finally \eqref{eq:before_i,j} and \eqref{eq:i,j} gives
\begin{align}
    p_{5,1} & \coloneqq \left(-1+1/w_p\right)\tilde{s}_{-\frac{1}{2},0} - (1/w_p) \tilde{s}_{\frac{1}{2},0}  &
    p_{5,2} & \coloneqq \left(-1+1/w_p\right) \tilde{s}_{0,-\frac{1}{2}} - (1/w_p) \tilde{s}_{0,\frac{1}{2}}     \\
    p_{5,3} & \coloneqq -(1/w_p)\tilde{s}_{-\frac{1}{2},0}                                                 &
    p_{5,4} & \coloneqq -(1/w_p) \tilde{s}_{0,-\frac{1}{2}}                                               \\
    p_{5,5} & \coloneqq - 1/\left(w_p(\lambda + \mu)\right)										& &	\\
    q_{5,1} & \coloneqq \left(1-1/w_p\right)\tilde{s}_{\frac{1}{2},0}                                &
    q_{5,2} & \coloneqq \left(1-1/w_p\right)\tilde{s}_{0,\frac{1}{2}}                                  \\
    q_{5,3} & \coloneqq \left(1-1/w_p\right)\left(1/(\lambda + \mu)\right) .
\end{align}
These coefficients form the desired matrices
\begin{equation} \label{eq:PandQ}
    P(\theta) \coloneqq
    \begin{pmatrix}
        p_{1,1} & 0       & p_{1,3} & 0       & p_{1,5} \\
        p_{2,1} & 0       & p_{2,3} & 0       & p_{2,5} \\
        0       & p_{3,2} & 0       & p_{3,4} & p_{3,5} \\
        0       & p_{4,2} & 0       & 0       & p_{4,5} \\
        p_{5,1} & p_{5,2} & p_{5,3} & p_{5,4} & p_{5,5}
    \end{pmatrix}
    \quad \text{and} \quad
    Q(\theta) \coloneqq
    \begin{pmatrix}
        q_{1,1} & 0       & q_{1,3} \\
        q_{2,3} & 0       & q_{2,3} \\
        0       & q_{3,2} & q_{3,3} \\
        0       & q_{4,2} & q_{4,3} \\
        q_{5,1} & q_{5,2} & q_{5,3}
    \end{pmatrix}.
\end{equation}
Now, we can invert $P$ and take $\tilde{S}_h$ to be the lower $3\times 3$ block of $P^{-1}Q$, as described in~\eqref{eq:Sh_def}.

\begin{figure}
\begin{center}
	\newcommand{\image}[1]{\includegraphics[width=0.24\linewidth]{./images/#1}}
    \subfigure[\footnotesize Damping]{\image{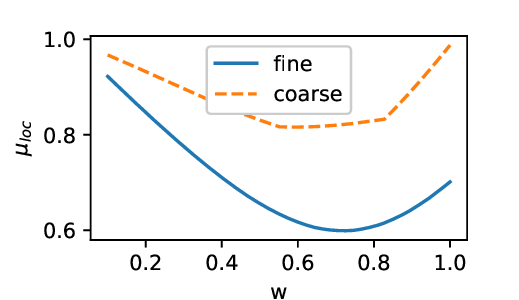}\label{fig:muloc_vs_w}} 
    \subfigure[\footnotesize Frequency]{\image{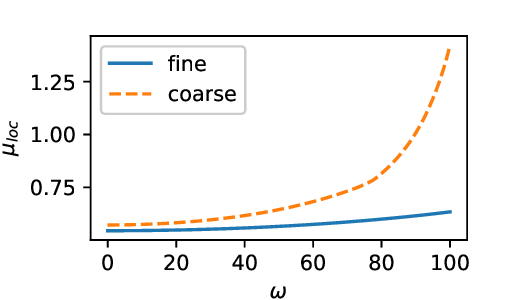}\label{fig:muloc_vs_kappa}}
    \subfigure[\footnotesize Poisson's ratio]{\image{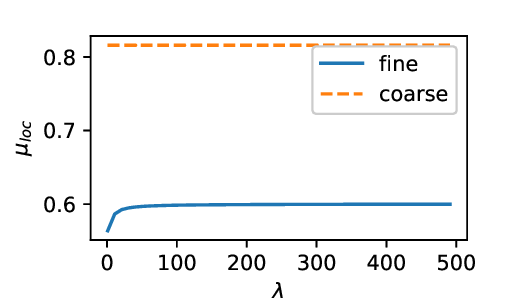}\label{fig:muloc_vs_lambda}}
    \subfigure[\footnotesize Attenuation]{\image{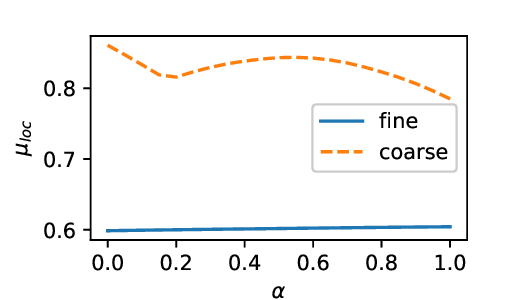}\label{fig:muloc_vs_gamma}}\\
\end{center}
   \caption{The influence of the damping coefficient $w$, the Poisson's ratio through the Lam{\'e} coefficient $\lambda$, the frequency $\omega$ and the attenuation parameter $\gamma$ on the smoothing factor $\mu_{loc}.$
      (a) $\mu_{loc}$ vs. $w$, for $\omega=80$, $\gamma = 0.2,\,\lambda = 500$ and $\mu = 1$.
     (b) $\mu_{loc}$ vs. $\lambda$, with $w_h=0.75$ for the fine grid and $w_H=0.6$ for the coarse grid, $\omega=80$, $\gamma = 0.2$ and $\mu = 1$.
            (c) $\mu_{loc}$ vs. $\omega$, with $w_h=0.75$ for the fine grid and $w_H=0.6$ for the coarse grid, $\gamma = 0.2,$  $\lambda = 500$ and $\mu = 1$.
       (d) $\mu_{loc}$ vs. $\gamma$, with $w_h=0.75$ for the fine grid and $w_H=0.6$ for the coarse grid, $\omega=80,$ $\lambda = 500$ and $\mu = 1$.} 
\label{Fig_muloc_all}
\end{figure}

\subsection{Numerical examples}

We fix a fine grid $h=1/128$ and a coarse grid $H=1/64.$ For both grids, we fix the density $\rho = 1$ and vary other parameters. We sample $63\times 63$ equally spaced frequencies $\theta = (\theta_1,\theta_2) \in [-\pi/2,3\pi/2]^2$  and calculate the spectral radius of $\tilde{S}_h(\theta)$. Taking the maximum over $\theta\in T^{high}$ we estimate the smoothing factor $\mu_{loc}.$

The choice of a wavenumber $\omega$ is done by a well known rule-of-thumb:  the largest wavenumber that can be represented on a given grid satisfies $\omega h \leq \pi/5,$ see e.g.~\cite{elman2001multigrid}. Thus, we take $\omega = 80$ as the highest frequency of a well represented wave on our fine grid. We can see in Fig.~\ref{fig:muloc_vs_kappa} that for the coarse grid, things get worse earlier. However, $\omega = 80$ gives a reasonable smoothing factor for both grids. The attenuation has essentially no effect on the smoothing, see Fig.~\ref{fig:muloc_vs_gamma}. This is not surprising, since the role of the attenuation is more related to the coarse grid correction, which we do not analyze here. We fix $\gamma = 0.2$, which has been observed as a reasonable attenuation in practice. Finally, we choose Lam{\'e} parameters $\lambda = 500$ and $\mu = 1,$ that yield a Poisson's ratio of $0.499$ to illustrate that we have a good scaling w.r.t. the Poisson's ratio. That is, as seen in Fig.~\ref{fig:muloc_vs_lambda}, for both the coarse and the fine grid we see almost no influence of $\lambda$ on the smoothing factor.

We first search for an optimal damping parameter assuming the same damping parameter for all components $w=w_u=w_v=w_p$. With the above parameters fixed, we let the damping vary between $0$ and $1$ and observe in Fig.~\ref{fig:muloc_vs_w} that the optimal damping parameters are approximately $w_h \approx 0.73$ for the fine grid and $w_H \approx 0.6$ for the coarse grid. These theoretical values nearly coincide with the optimal damping parameters we see in practice, of $0.75$ and $0.5$ respectively, which we use in our 2D experiments, see Section \ref{sec:NumericalResults}. 

\begin{figure}
\begin{center}
	\newcommand{\image}[1]{\includegraphics[width=0.4\linewidth]{./images/#1}}
    \subfigure[\footnotesize $\rho_{loc}$ for damping combinations]{\image{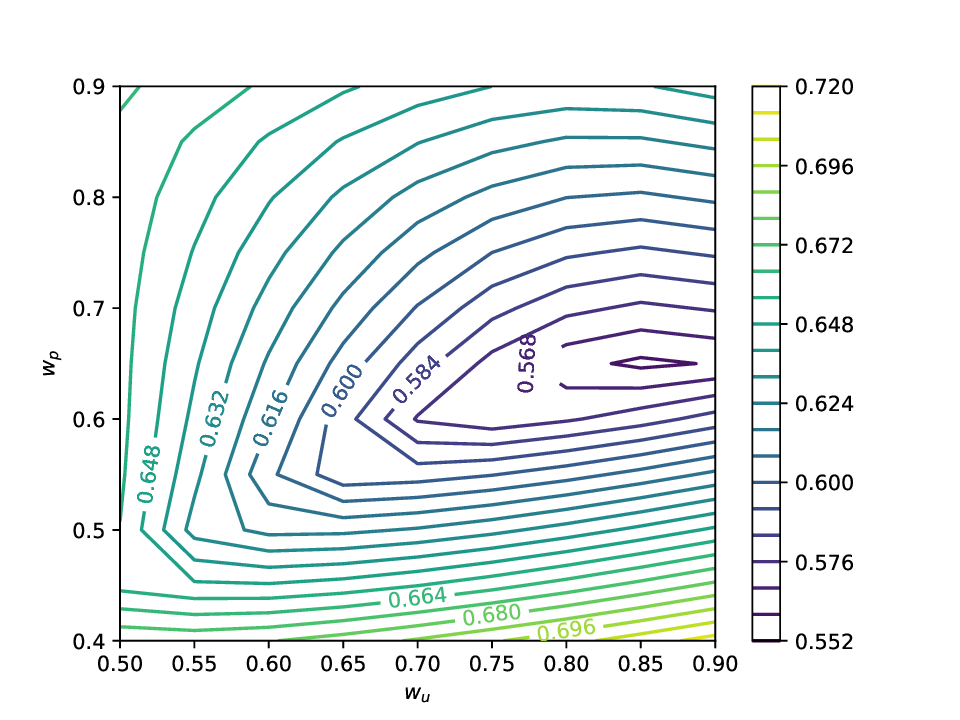}\label{fig:muloc_vs_w_2d}} 
    \subfigure[\footnotesize Amplification factor]{\image{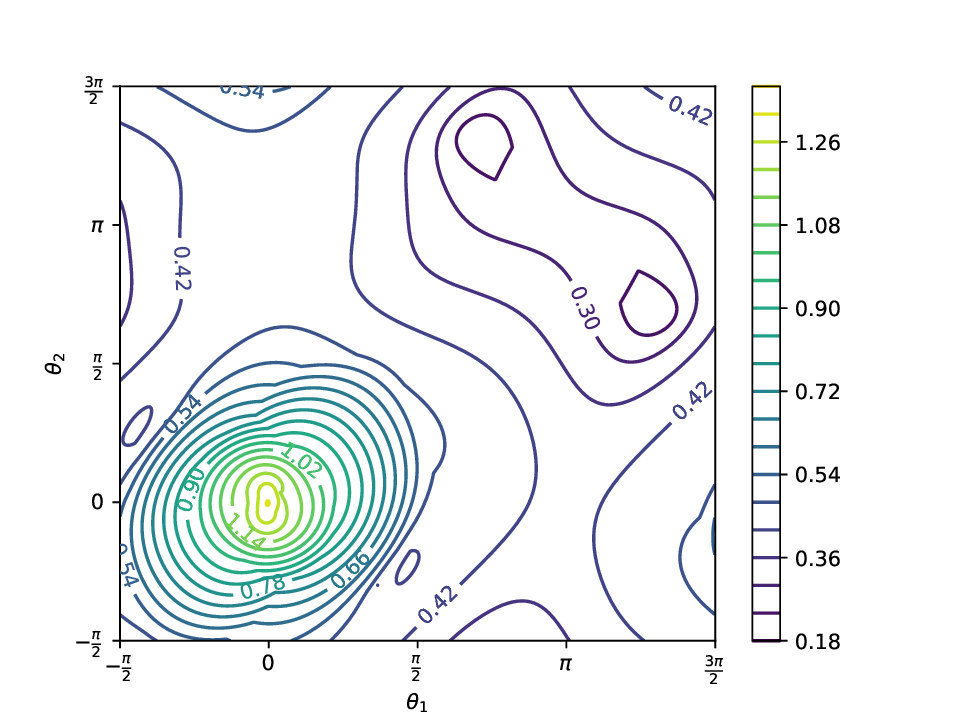}\label{fig:amplificationContour}}
    \\
\end{center}
   \caption{On the left, the two-grid factor $\rho_{loc}$ as a function of $w_u$ and $w_p$ for $h=1/128$, frequency $\omega=80$, attenuation $\gamma = 0.2$ and Lam{\'e} coefficients $\lambda = 500, \mu = 1$ and density $\rho=1$. On the right, amplification factor as a function of $-\pi/2\leq \theta\leq 3\pi/2$ with the same parameters, and the optimal damping parameters found on the left, $w_u=0.85$ and $w_p=0.65$.} 
\label{fig:contour}
\end{figure}

We apply a similar experiment, allowing the damping parameters $w_u$ and $w_p$ be different, but demanding that $w_u=w_v$. Optimizing over $w_u,w_p\in[0.5,0.9]$ in jumps of $0.05$, we observe numerically that $w_u=0.85$ and $w_p=0.65$ are optimal parameters, see Figure \ref{fig:muloc_vs_w_2d}. The corresponding numerically calculated smoothing factor for one damping and for component-dependant damping, on the fine grid, is
\begin{equation} \label{eq:muloc_results}
    \mu_{loc}(w) = 0.58 \quad \text{and} \quad \mu_{loc}(w_u,w_p) = 0.55
\end{equation}
This value is obtained by taking maximum over high frequencies for the amplification factor. For the case of component-dependent damping, the amplification as a function of the frequency is depicted in Fig.~\ref{fig:amplificationContour}. It is shown to have a typical smoothing behavior: it damps the high frequencies while almost not interfering with the low ones. This shows that the economic Vanka smoother is suitable for the elastic Helmholtz equation in mixed formulation.

Next, we present two-grid results, and compare them to the convergence rate of the multigrid cycle in practice. For that purpose, we define the convergence factor as
\begin{equation} \label{eq:convFactor}
c_f^{(k)} = \left(\frac{\| r_k \|}{\| r_0 \|}\right)^{1/k}
\end{equation}
where $r_0$ denotes the residual after a warm-up of 5 iterations, for the error-residual equation of $TG$, and $r_k$ denotes the residual after $k$ more iterations. We take $k$ to be the smallest number of iterations such that $r_k<10^{-9}$. 

In Table \ref{tab:rho_cf_comparison} we compare the multigrid performance and the convergence in practice. For this goal, we measure $c_f$ from \eqref{eq:convFactor} for a two-grid cycle with 1 pre- and 1 post-relaxation, with lexicographic ordered economic Vanka as a smoother (the same as in our analysis). We compare it to the two-grid factor $\rho_{loc}$ from Definition \ref{def_rho_loc}, and as a reference value (represents the case of an ideal coarse grid correction) we compare to $\mu_{loc}^2$, with the smoothing factor $\mu_{loc}$ from Definition \ref{def_mu_loc}. We observe that, especially for higher grid points per wavelength or lower attenuations, the two-grid factor gives a more reliable prediction.

\begin{table}
\centering
\begin{tabular}{c|cc|cc|cc}
\toprule
  \mc{7}{c}{LFA vs. convergence in practice}\\
 \midrule
  & \mc{2}{c|}{10 grid points} & \mc{2}{c|}{8 grid points} & \mc{2}{c}{6.6 grid points} \\
    & $\gamma=0.1$ & $\gamma=0.15$ & $\gamma=0.15$ & $\gamma=0.2$ & $\gamma=0.2$ & $\gamma=0.3$ \\
\midrule
$c_f$         & 0.69 & 0.51 & 0.71 & 0.64 & 0.77 & 0.61 \\
$\rho_{loc}$  & 0.79 & 0.56 & 0.79 & 0.61 & 0.82 & 0.56 \\ 
$\mu_{loc}^2$ & 0.35 & 0.35 & 0.38 & 0.38 & 0.44 & 0.44 \\ 
\bottomrule
 \end{tabular}
\caption{The LFA two-grid factor $\rho_{loc}$ and the convergence factor in practice $c_f$ for a grid of $h=1/1024$ with $\lambda=500$, $\mu=1$ and $\rho=1$ with damping of $w=0.75$. As a reference value, $\mu_{loc}^2$ resembles a two-grid with 1 pre- and 1 post-smoothing, assuming an ideal coarse grid correction. The calculation of $c_f$ is done with a Vanka smoother in a lexicographic order.}
\label{tab:rho_cf_comparison}
\end{table}

\section{Numerical results}\label{sec:NumericalResults}

In this section we demonstrate the hybrid shifted Laplacian multigrid and domain decomposition preconditioner for the elastic Helmholtz equation. We present examples that appear in geophysical applications, where typically the length of the domain is quite high (about 20km) compared to its depth (about 5km). We first perform a few experimental comparisons in 2D, and then perform 3D experiments. Note that the 2D experiments are done only for illustrating the behavior of the solver for 3D problems at those scales. Our code is written in the {\tt Julia} language \cite{Julia}, and is part of the {\tt jInv.jl} package \cite{jInv17}. Using this package, our code can be easily used as a forward solver for three-dimensional elastic full waveform inversion in the frequency domain.

In subsection \ref{subsec:ResultsShifted} we demonstrate the performance of the shifted Laplacian solver applied to the elastic Helmholtz equation. In the first experiment we observe that choosing the mixed formulation, together with a choice of cell-wise relaxation, gives results that are scalable w.r.t to Poisson ratio. We compare these results to applying shifted Laplacian multigrid to the original formulation of the elastic equation with Jacobi as a smoother, which does not have this scaling property. In the second experiment we show that our results for the elastic equation using mixed formulation are comparable to shifted Laplacian multigrid applied to the acoustic equation using the shear velocity.

In subsection \ref{subsec:ResultsDD} we apply the combination of multigrid and DD. In the first experiment we demonstrate the performance of DD alone. In the second experiment we apply DD as a coarse grid solver, and in the third experiment we apply DD in the fine level, with or without applying it in the coarse grid within the multigrid cycle. Applying DD in the fine grid, before applying the multigrid cycle, allows a trivial parallelism that is highly desired here because of the size of the problem. Applying DD in the coarse grid allows to tackle the relatively large coarse grids needed to represent the high wavenumber. Our experiments approve that on both ends of the multigrid cycle, the convergence rate of the hybrid method with a moderate number of subdomains is comparable to the convergence rate of each of the methods alone.

In all the examples we use the preconditioned GMRES(5) Krylov solver \cite{saad1993flexible}, and seek a solution with relative residual accuracy of $10^{-6}$, starting from a zero initial guess. The right hand side $\vec\bfq$ is chosen to be a point source located at middle of the top row of the domain, similarly to Fig. \ref{fig:wavefields}. Unless stated otherwise, we use the shifted operator \eqref{eq:shift} as preconditioner, using a shift parameter of 0.1, 0.3, and 0.4 for two-, three- and four-level MG methods, respectively. 
In all the experiments that concern the solution of the reformulated system \eqref{eq:elasticHelmReformulatedSystem}, we use $W(1,1)$-cycles using red-black cell-wise relaxation, with a damping parameter of 0.75, 0.5, and 0.25 on the first, second and third grids. While our theoretical analysis in Section \ref{sec:LFA} is valid for the lexicographic version only, our choice of a constant damping parameter for red-black Vanka in the next subsections is based on trial and error. The choice for the coarse grids is motivated by \cite{calandra2013improved} which used a smaller damping parameters on coarse grids for the acoustic \eqref{eq:acousitcHelm}. In the multigrid framework, we define our coarse grid problems by the Galerkin product \eqref{eq:coarseGridMatrix}. These are our ``default'' parameters that we found to be the most effective. In all the examples, we use an absorbing boundary layer of 20 cells, and add a small artificial attenuation by adding $0.01\pi$ to $\gamma$ in \eqref{eq:elasticHelmMass}. The code for running the experiments below is available online at
\url{https://github.com/JuliaInv/Helmholtz.jl}
together with
\url{https://github.com/JuliaInv/Multigrid.jl}
as modules that are part of the {\tt jInv.jl} package \cite{jInv17}.

\subsection{Shifted Laplacian multigrid for the elastic Helmholtz equation} \label{subsec:ResultsShifted}

In our first experiment we use a two-dimensional constant coefficients example, and monitor the influence of $\bflambda$ and $\omega$ on the performance of the shifted Laplacian method, while keeping $\bfrho$ and $\bfmu$ fixed. The case of nearly incompressible material correspond to a case where $\lambda \gg \mu$, or when the Poisson's ratio
$$
\sigma = \frac{\lambda}{2(\lambda+\mu)}
$$
tends to 0.5. On the one hand we show a standard shifted Laplacian approach without using the mixed formulation, applying W(2,2)-cycles with damped Jacobi relaxation as preconditioner. On the other hand, we show the shifted Laplacian approach using the mixed formulation \eqref{eq:elasticHelmReformulatedSystem} using W(1,1)-cycles with red-black cell-wise relaxation. Fig. \ref{fig:CompTime} summarizes the results. It is clear that the standard shifted Laplacian method performs reasonably well in cases when $\bflambda \approx \bfmu$. Similarly to the acoustic Helmholtz, difficulties are observed when the frequency is high. This behavior is expected. However, the standard method deteriorates as $\bflambda$ gets larger ($\bfmu$ is fixed), while the proposed approach using the mixed formulation \eqref{eq:elasticHelmReformulatedSystem} and cell-wise smoothing, does not deteriorate at all as $\bflambda$ grows. Although the results are slightly worse for economy Vanka compared to full Vanka as a smoother,
it requires less FLOPs, as mentioned in Section \ref{sec:Method}. However, qualitatively both show the same scaling. This scaling is exactly what our method aims to achieve.

\begin{figure*}[t]
	\begin{center}
		\newcommand{\rottext}[1]{\rotatebox{90}{\hbox to 30mm{\hss  #1\hss}}}
		\iwidth=40mm
		\iheight=40mm
		\begin{tabular}{@{}c@{}c@{}c@{}c@{}}
		\rottext{\hspace{50pt} $\#$ iterations}   & 



\begin{tikzpicture}
\begin{axis}[%
width=\iwidth,
height=\iheight,
at={(0,0)},
scale only axis,
log ticks with fixed point,
xmin=0.4,
xmax=20,
xtick={0.5, 1,2,4,8,16},
ytick={25,50,100,200,400},
ymin=20,
ymax=500,
ymode=log,
xmode=log,
xticklabels from table={./images/lambdaAxis.dat}{input},
xlabel = $\lambda$,
legend style={font=\footnotesize,
legend pos = north west},
]
]
\addplot [color=black,thick,solid,mark=*]
  table[row sep=crcr]{%
0.5 93 \\
1   105  \\
2   131  \\
4   180 \\
8   268 \\
16   442 \\
};
\addplot [color=blue,thick,solid,mark=square*]
table[row sep=crcr]{%
0.5 50 \\
1   50  \\
2   54  \\
4   55 \\
8   56 \\
16   57 \\
};
\addplot [color=red,thick,solid,mark=square*]
table[row sep=crcr]{%
0.5 41 \\
1   41  \\
2   42  \\
4   42 \\
8   42 \\
16  42 \\
};


\end{axis}
\end{tikzpicture}%
 &  \hspace{8pt}



\begin{tikzpicture}
\begin{axis}[%
width=\iwidth,
height=\iheight,
at={(0,0)},
scale only axis,
log ticks with fixed point,
xmin=0.4,
xmax=20,
xtick={0.5, 1,2,4,8,16},
ytick={25,50,100,200,400},
ymin=20,
ymax=500,
ymode=log,
xmode=log,
xticklabels from table={./images/lambdaAxis.dat}{input},
xlabel = $\lambda$,
]
\addplot [color=black,thick,solid,mark=*]
  table[row sep=crcr]{%
0.5 217 \\
1   272  \\
2   386  \\
20   1500 \\
};
\addplot [color=blue,thick,solid,mark=square*]
table[row sep=crcr]{%
0.5 104\\
1   107  \\
2   110  \\
4   112 \\
8   113 \\
16  110 \\
};
\addplot [color=red,thick,solid,mark=square*]
table[row sep=crcr]{%
0.5 76\\
1   76  \\
2   78  \\
4   80 \\
8   78 \\
16  75 \\
};


\end{axis}
\end{tikzpicture}%
& \hspace{2pt} 

\begin{tikzpicture}
\begin{axis}[%
width=\iwidth,
height=\iheight,
at={(0,0)},
scale only axis,
log ticks with fixed point,
xmin=0.4,
xmax=20,
xtick={0.5, 1,2,4,8,16},
ytick={25,50,100,200,400},
ymin=20,
ymax=500,
ymode=log,
xmode=log,
xticklabels from table={./images/lambdaAxis.dat}{input},
xlabel = $\lambda$,
legend style={font=\footnotesize,
legend pos = south east},
]
\addplot [color=black,thick,solid,mark=*]
  table[row sep=crcr]{%
0.5 1500 \\
1   1500 \\
2   1500  \\
20   1500 \\
};
\addplot [color=blue,thick,solid,mark=square*]
table[row sep=crcr]{%
0.5 267\\
1   268  \\
2   284  \\
4   281 \\
8   281 \\
16  266 \\
};
\addplot [color=red,thick,solid,mark=square*]
table[row sep=crcr]{%
0.5 181\\
1   183  \\
2   187  \\
4   202 \\
8   182 \\
16  186 \\
};


\addlegendentry{\scriptsize Standard, $\omega$-Jac}
\addlegendentry{\scriptsize Mixed, EconVanka}
\addlegendentry{\scriptsize Mixed, FullVanka}

\end{axis}
\end{tikzpicture}%
\\
		&\hspace{20pt} (a) $n_{cells} = 256\times128,\omega=3\pi$  &\hspace{20pt}  (b) $n_{cells} = 512\times256,\omega=6\pi$ &\hspace{20pt} (c) $n_{cells} = 1024\times512,\omega=12\pi$\\
		\end{tabular}
	\end{center}
	\caption{Number of preconditioning cycles needed for convergence with shifted Laplacian multigrid for 2D elastic Helmholtz, using standard vs. mixed formulation, where the latter is used with full and economic red-black Vanka smoothers. We use constant coefficients: $\bfmu=\bfrho = 1$, and change $\bflambda$ and $\omega$. The highest frequency corresponds to 10 grid points per shear wavelength. The value $\lambda=16$ corresponds to a Poisson's ratio of $0.47$.}
	\label{fig:CompTime}
\end{figure*}

In Table \ref{tab:RB_comparison} we compare the performance of our preconditioner with different choices of Vanka smoothers. We compare red-black and lexicographical ordered multiplicative Vanka as well as additive Vanka. We use constant relaxation damping $w$ of 0.75 and 0.5 for the two grids, respectively, and component-dependant relaxation damping: $(w_u,w_p)$ is $(0.85,0.65)$ and $(0.6,0.4)$ for the two grids, respectively. We observe that the additive Vanka has the worst performance, whereas the different orderings of multiplicative Vanka has a somewhat similar and better performance. The slightly better results for lexicographic order can be explained by the choice of damping parameters, that are optimal for lexicographic order according to the analysis. For the sake of parallelism, in the rest of the numerical results we use red-black ordered multiplicative Vanka. Although the performance of component-dependant damping is slightly better than constant damping, the improvement is not significant. For the sake of simplicity, in the rest of the results we use constant damping.

\begin{table}
\centering
\begin{tabular}{c|cc|cc|cc}
\toprule
  \mc{7}{c}{Comparison of different versions of Vanka smoothers}\\
 \midrule
  & \mc{2}{c|}{$n_{cells} = 256\times128,\omega=3\pi$} & \mc{2}{c|}{$n_{cells} = 512\times256,\omega=6\pi$} & \mc{2}{c}{$n_{cells} = 1024\times512,\omega=12\pi$} \\
  & $w_u=0.75$ & $w_u=0.85$ & $w_u=0.75$ & $w_u=0.85$ & $w_u=0.75$ & $w_u=0.85$ \\
    & $w_p=0.75$ & $w_p=0.65$ & $w_p=0.75$ & $w_p=0.65$ & $w_p=0.75$ & $w_p=0.65$ \\
\midrule
Red-black     & 42 & 38 &  84  & 75 & 186 & 169\\
Lexicographic & 36 & 34 &  75  & 65 & 152 & 137\\
Additive      &  60&  55 & 124 & 108 & 335 & 280\\
\bottomrule
 \end{tabular}
\caption{The number of preconditioning cycles needed for convergence with shifted Laplacian multigrid for 2D elastic Helmholtz, using different versions of Vanka smoothers with different damping approaches. We use constant coefficients: $\bfmu=\bfrho = 1$, and $\bflambda=16$. We choose $\omega$ in correspondence with the grid size. The frequency corresponds to 10 grid points per shear wavelength. The value $\lambda=16$ corresponds to a Poisson's ratio of $0.47$.}
\label{tab:RB_comparison}
\end{table}

\paragraph{Shifted Laplacian Multigrid for the Elastic vs. Acoustic Helmholtz Equations}

\begin{figure}
\centering
\includegraphics[width=0.9\textwidth]{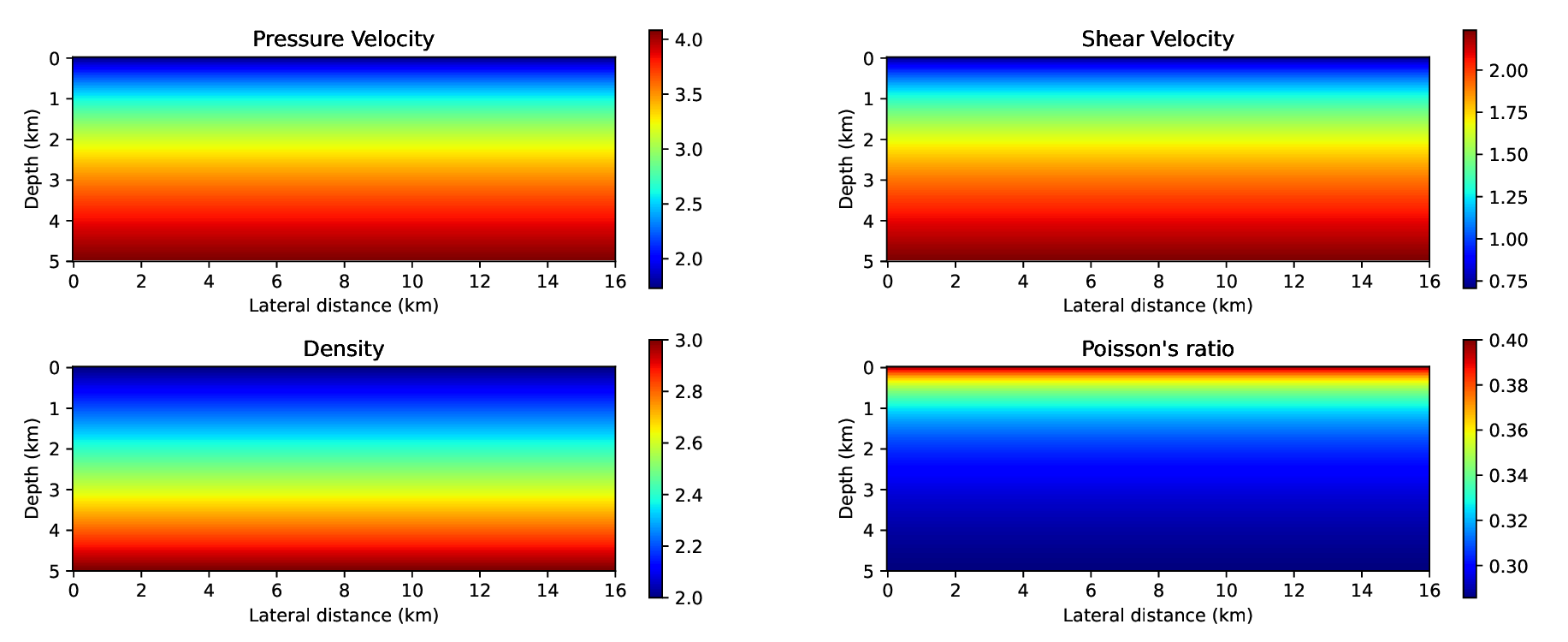}
\caption{The elastic linear model. Velocity units are in $km/sec$. }
\label{fig:LinearVel}
\end{figure}

Our second set of experiments aims to demonstrate one of our main messages in this work: the new shifted Laplacian method solves the elastic \eqref{eq:elasticHelm} with approximately the same performance of the standard shifted Laplacian for the acoustic equation \eqref{eq:acousitcHelm}, only with respect to the shear wavenumber instead of the lower pressure wavenumber. To demonstrate this we use the linear velocity and density models depicted in Fig. \ref{fig:LinearVel}.

\begin{table} 
\centering
\begin{tabular}{c|cc|cc|cc}
\hline
  \toprule
  \mc{7}{c}{Shifted Laplacian: acoustic (shear) vs. elastic}\\
  \midrule
    Grid size  &  \mc{2}{c|}{$400\times128$} &  \mc{2}{c|}{$800\times256$} & \mc{2}{c}{$1600\times512$} \\
$\omega$ & $2.4\pi$ &   $3.5\pi$ &  $4.7\pi$  & $7.1\pi$ & $9.4\pi$ & $14.2\pi$ \\
  \midrule
Acoustic & 25 & 40 & 45 & 86 & 75 & 196 \\
Elastic & 27 & 37 & 47 & 78 & 79 & 148 \\
  \bottomrule
 \end{tabular}
\caption{Number of preconditioning cycles needed for convergence with acoustic and elastic shifted Laplacian multigrid for the 2D linear model presented in Fig \eqref{fig:LinearVel}. For the acoustic equation we use damped Jacobi with damping 0.8 as a smoother, and for the elastic equation with mixed formulation we use red-black cell-wise full Vanka smoother with damping 0.75 and 0.5 on the first and second grids, respectively.}
\label{tab:AcousticComp}
\end{table}

We solve both the acoustic and elastic equations for three grid-sizes using 10 and 15 grid-points per wavelength with respect to the shear velocity. The acoustic equation is defined with the shear velocity instead of the pressure velocity. Both equations are solved with 3-level W-cycles, and a shift parameter $\alpha = 0.2$. The acoustic equation is solved by $W(2,2)$-cycles with damped Jacobi as  a smoother (with damping of 0.8) as in \cite{JointEikFWI17}. For the elastic equation we use full red-black Vanka as a smoother (one pre- and one post-smoothing). Table \ref{tab:AcousticComp} summarizes the results. It is clear that the performance of the shifted Laplacian in the two scenarios is comparable, even though the problems are different, and the discretization and relaxation methods are different. This shows that the ability to solve the acoustic equation is the key to solve the elastic equation. The main difficulty of the problem lies only in the indefiniteness of the linear systems to be solved.

\subsection{Combination of multigrid with domain decomposition} \label{subsec:ResultsDD}

\begin{figure}
\centering
\includegraphics[width=1.0\textwidth]{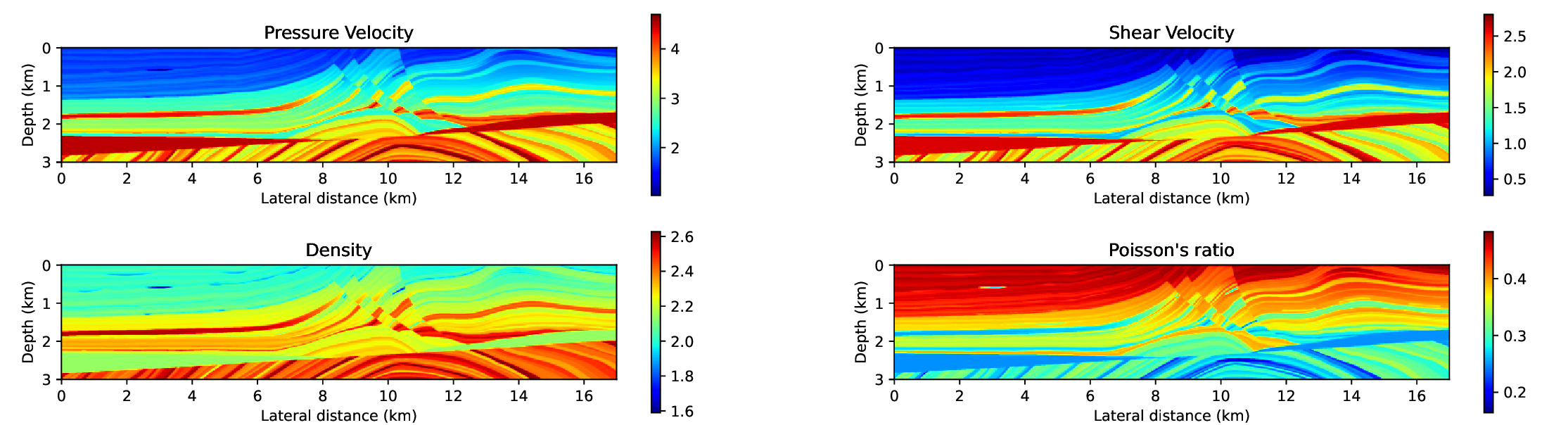}
\caption{The Marmousi2 elastic model. Velocity units are in $km/sec$. }
\label{fig:Marmousi2}
\end{figure}

In this subsection we demonstrate the performance of our hybrid solver on the highly heterogeneous Marmousi2 elastic two-dimensional model \cite{martin2006marmousi2}, which appears in Fig.~\ref{fig:Marmousi2}. This is a 2D model, but we consider it as a case study for real 3D scenarios. Since the model is shallow (3 km deep), we extend it by half a km at the bottom to accommodate the absorbing boundary layer.  In all the 2D experiments, we choose the frequency $\omega$ to correspond to about 12 grid-points per shear wavelength.

In the first experiment we solve the elastic equation using the DD method only, using multi-colored scheduling for the solution of the subdomains, as demonstrated in Fig. \ref{fig:DDcolors}. We report our results in Table~\ref{tab:DDexact}, where we either use Dirichlet interface conditions for the shifted equation, or absorbing boundary conditions. We observe that when the domain is divided aggressively, the convergence of the DD method deteriorates. It settles down with the local behaviour of the system: too small subdomains cannot contain the attenuated wave. We also see that absorbing boundary conditions perform better than Dirichlet boundary conditions with added attenuation. The absorbing boundary conditions take care of reflections from the interfaces, and hence allow to sub-divide the domain further, even when the attenuation is small, and the wave propagation takes a large region in the subdomain.

\begin{table}
\centering
\setlength{\tabcolsep}{3pt}
\begin{tabular}{cc|ccccc|ccccc}
\toprule
  & & \mc{5}{c|}{DD(Dir. $\alpha=0.2$).} & \mc{5}{c}{DD(ABC, $\alpha=0.05$).}\\
  Grid size &$\omega$  &  $2\times 1$ &$4\times 1$ & $8\times 1$ &$16\times 1 $ & $16\times 2 $ & $2\times 1$ &$4\times 1$ & $8\times 1$ & $16\times 1$ & $16\times 2$\\
  \midrule
$544\times112$ & $1.8\pi$  &$59$& 63 &    85 & 290 &  380$^*$  & 28 & 35 & 54 & 130 & 171$^*$ \\
$1088\times224$ & $3.6\pi$ &131& 140 &  $160$ & 264 &  393$^*$ & 48  & 59 & 80 & 162 & 263$^*$ \\
$2176\times448$& $7.0\pi$ &  260& 270 & 298 & 396 & $>500$  & 77 & 90 & 114 & 175 & 264 \\
  \bottomrule
 \end{tabular}
\caption{Number of preconditioning cycles needed for convergence for the 2D elastic Marmousi2 model, when an exact solution is obtained for the subdomains. $*$ marks cases where the shift had to be increased by 0.05 compared to what is written, because the number of subdomains was large compared to the grid size.}
\label{tab:DDexact}
\end{table}

In the second experiment we apply our shifted Laplacian method to the elastic equation with mixed formulation and compare between using an exact solver on the coarsest grid and using DD as a coarse grid solver. The results are summarized in Table~\ref{tab:MG}. In the left three columns we see the performance of the multigrid preconditioner for 2,3 and 4 levels.

By comparing the left columns with the right columns of Table~\ref{tab:MG}, we observe that adding DD in the coarse grid does not hampers the convergence as long as the division is not too aggressive. That is, at the worst, we observe some deterioration from the MG-DD two-level method. That is because the two-level MG method relies mostly on the coarsest-grid solution and is less local (less attenuated), hence, adding DD on the coarsest-level hurts convergence. On the other hand, using 3 and 4 levels, we hardly see any deterioration. In any case, we show that it is possible to choose a proper size of subdomains, such that the results for the hybrid method are similar to the results for the MG method with exact coarse grid solution.

\begin{table}
\centering
\begin{tabular}{cc|ccc|ccc}
\toprule
  & & \mc{3}{c|}{MG-exact} & \mc{3}{c}{MG-DD}\\
  Grid size &$\omega$  &  $^{lev=2}_{\alpha=0.1}$ & $^{lev=3}_{\alpha=0.2}$ & $^{lev=4}_{\alpha=0.4}$ & $^{lev=2,\alpha=0.1}_{dom=(2,4,8)\times1}$ & $^{lev=3,\alpha=0.2}_{dom=(2,4,8)\times1}$ & $^{lev=4,\alpha=0.4}_{dom=(2,4,8)\times1}$\\
  \midrule
$544\times112$ & $1.8\pi$  & 29   & 47    & 81  & (36, 39, 50) &  (47, 49, 50) & (81, 81, 81)\\
$1088\times224$ & $3.6\pi$ & 54   & 97    & 172  & (73, 75, 89) & (99, 100, 101)  & (172, 172, 172)\\
$2176\times448$& $7.0\pi$  & 108   & 220    & 379 & (139, 154, 163) & (226, 229, 230) & (379, 379, 379) \\
  \bottomrule
 \end{tabular}
\caption{Number of preconditioning cycles needed for convergence for the 2D elastic Marmousi2 model. For all the MG-DD results, the subdomains on the coarsest grid were chosen to be $34\times 14$, and the number of subdomains was chosen according to the coarsest grid size (the overlap is 2 cells), so that all the MG-DD results scale linearly in memory and computation per MG cycle.}
\label{tab:MG}
\end{table}

\begin{table}
\centering
\begin{tabular}{cc|cccccccc}
\toprule
  & & \mc{6}{c}{DD(ABC)-MG-exact} \\
  Grid size &$\omega$  & $^{dom=2\times 1}_{lev=2, \alpha=0.1}$  & $^{dom=4\times1}_{lev=2, \alpha=0.1}$ & $^{dom=8\times 1}_{lev=2, \alpha=0.1}$ & $^{dom=2\times 1}_{lev=3, \alpha=0.2}$ & $^{dom=4\times1}_{lev=3,\alpha=0.2}$  & $^{dom=8\times 1}_{lev=3, \alpha=0.2}$ \\
\midrule
$544\times112$ & $1.8\pi$  &    37 &  42 &  53 &  57 &  62 &  68 \\
$1088\times224$ & $3.6\pi$ &    69  & 77 &  93 & 115 & 124 & 137 \\
$2176\times448$& $7.0\pi$  &    130  & 133 & 147 & 250 & 254 & 262\\
\midrule
  & & \mc{6}{c}{DD(ABC)-MG-DD ($coarse\;domains=2\times1$)}\\
\midrule
$544\times112$ & $1.8\pi$  & 43 &  54 & 114 &  57 &  65 &  71 \\
$1088\times224$ & $3.6\pi$ &  79 &  94 & 158 & 115 & 125 & 139 \\
$2176\times448$& $7.0\pi$  & 153 &   180 &  213 &  254 &  254 & 257 \\
\midrule
  & & \mc{6}{c}{DD(ABC)-MG-DD ($coarse\;domains=4\times1$)}\\
\midrule
$544\times112$ & $1.8\pi$  &   51 & 113 & 190 &  58 &  76 & 100  \\
$1088\times224$ & $3.6\pi$ &   90 & 154 & $>$500 & 118 & 130 & 459 \\
$2176\times448$& $7.0\pi$  &    169 & 219 & 467 & 252 & 265 & 327 \\

\bottomrule

 \end{tabular}
\caption{Number of preconditioning cycles needed for convergence for the 2D elastic Marmousi2 model. Top-level uses ABC at the interfaces.}
\label{tab:hybridMGDD}
\end{table}

To reduce the size of the coarsest grid, we inevitably use 4 levels in our multigrid hierarchy. A similar effort was performed in \cite{calandra2013improved} for the acoustic equation, where the authors show that four levels can be used, but with a lower damping parameter for the relaxation on the third grid. We apply a similar strategy here, using a damping parameter of 0.5 for the Vanka relaxation on the first two grids, and 0.2 on the third. In addition, we also increase the shift parameter to $\alpha=0.4$, simply because the 4-level approach failed to converge for $\alpha = 0.2$. The rest of the parameters are the same as the default parameters mentioned earlier.

Finally, in Table~\ref{tab:hybridMGDD} we demonstrate the use of DD on the top (or outer) level. In the left columns we first divide the domain into subdomains, then apply the multigrid preconditioner within each domain where the coarsest grid solution is done exactly by an LU decomposition. The experiments in the left side are similar, only with DD as a coarse grid solver. Decomposing the domain before applying multigrid enhance distributed parallelism. Decomposing the domain (again) for solving the coarse problem, allows fewer multigrid levels. Compared to Table~\ref{tab:MG}, left and right side respectively, we see that adding DD outside the multigrid gives comparable performance and allows to enjoy parallelism without paying a meaningful price in the number of preconditioning cycles.

To summarize, using multigrid solely, we cannot use too many levels efficiently, and we end up with a quite large linear system for the coarsest grid in 3D. That is also a problem when solving the acoustic equation, only now the system is much larger (for the same mesh), especially when using the mixed formulation. Using DD alone, the domain must be divided aggressively. We demonstrate that the hybrid approach allows us to enjoy parallelism and use possibly fewer multigrid levels, without increasing the number of preconditioning cycles.

\subsection{Three-dimensional experiments}

\begin{figure}
\centering
\includegraphics[width=0.75\textwidth]{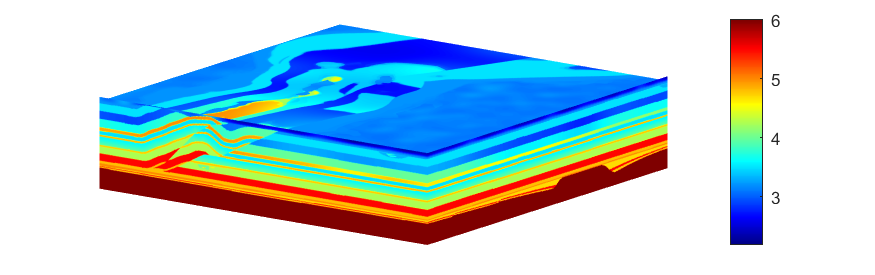}
\caption{The SEG Overthrust pressure velocity model ($V_p$). Units are in $km/sec$. The model corresponds to a domain of $20\times20\times4.65$ km.}
\label{fig:Overthrust}
\end{figure}

In this set of experiments, we demonstrate our ability to solve the problem in three dimensions. We use a three-dimensional version of the linear model in Fig. \ref{fig:LinearVel}, and in addition use the heterogeneous Overthrust model in Fig. \ref{fig:Overthrust}. Because the model is shallow, we add 16 grid points at the bottom of the domain to accommodate the absorbing boundary layer. Since this model is only acoustic and includes only pressure velocity, we set the shear velocity to be $V_s = 0.5V_p$ and set the density to be $\rho = 0.25V_p + 1.5$. The linear systems \eqref{eq:elasticHelmReformulatedSystem} are huge in 3D, and hence for these experiments we use much smaller grid sizes than in the 2D experiments. We apply our multigrid preconditioner using single precision computations to save memory. The tests were computed on a workstation with Intel Xeon Gold 5117 \@ 2GHz X 2 (14 cores per socket) with 256 GB RAM, running on Centos 7 Linux distribution.

Tables \ref{tab:hybridMGDD_3D_linear} and \ref{tab:hybridMGDD_3D_Overthrust} summarize our results in 3D. Note that the reported grid size is the number of cells in the domain, and the mixed formulation has four times the number of unknowns. Top-level subdomain sizes are considered as $128\times128$ or $64\times 64$ in the $x$-$y$ plane, while no division is applied in the $z$ direction. Three and four level multigrid are used with appropriate shift parameter and various number of domains on the coarsest grid. `oom' refers to the case where our machine ran out of memory, usually because the coarsest grid LU decompositions required too much memory, even in single precision (this happens for three levels).   

Generally, we see that the smaller the subdomains we use on the top-level, the more iterations are required. Moreover, when we use more levels in MG, we require more iterations as well. Interestingly, we see that splitting the domain decomposition between the top-level and coarsest grid (i.e., DD-MG-DD) yields better results than top-level decomposition alone (DD-MG) if we consider the same division overall for the coarsest grid. We see it in the tables when comparing the first vs. fifth columns, and the third vs. sixth columns. That is, using a less aggressive top-level subdomain division, and compensating for it by dividing the coarsest grid is beneficial. That it because in each W-cycle we visit the coarse grid multiple times, hence allow more information to go between the subdomains. We see that even a $4\times4\times1$ division on the coarsest grid (which would correspond to small $32\times32$ subdomains if the divisions were applied on the top-level only) do not lose much accuracy compared to the other counterparts.    

\begin{table}
\centering
\setlength{\tabcolsep}{3pt}
\begin{tabular}{cc|cccc|cc}
\toprule
\mc{2}{c|}{DD-MG-DD} & \mc{4}{c|}{{\textit{Top-level domain size in $x$-$y$: }$128\times128$}} &  \mc{2}{c}{$64\times64$} \\
  Grid size &$\omega$  & $^{lev=3, \alpha=0.2}_{coarse:\;2\times2\times 1}$  & $^{lev=3, \alpha=0.3^\dag}_{coarse:\;4\times 4 \times 1}$ & $^{lev=4, \alpha=0.4}_{coarse:\;2\times2\times 1}$ & $^{lev=4, \alpha=0.4}_{coarse:\;4\times4\times 1}$ &  $^{lev=3, \alpha=0.2}_{coarse:\;exact}$ &  $^{lev=4, \alpha=0.4}_{coarse:\;exact}$ \\
\midrule
$128\times128\times48$ & $ 1.13\pi$  & 18 & 25  & 34  & 34 & 20 & 34 \\
$192\times192\times72^*$ &  $ 1.69\pi$ & 25  & 34 &  55 & 55 & 28 & 50\\
$256\times256\times96$ & $ 2.26\pi$  & 38 & 54 & 75 & 75  & 40 & 68\\
$384\times384\times144$ & $3.38\pi$ & oom & oom & 145 & 146 & oom & 145 \\
\bottomrule
 \end{tabular}
\caption{Three-dimensional experiments of the hybrid solver for the linear model. $\dag$The attenuation was enlarged compared to 0.2 since the aggressive division of the coarsest grid solution led to inefficient solver. $^*$The top-level domain size is $192\times192$.}
\label{tab:hybridMGDD_3D_linear}
\end{table}

\begin{table}
\centering
\setlength{\tabcolsep}{3pt}
\begin{tabular}{cc|cccc|cc}
\toprule
\mc{2}{c|}{DD-MG-DD} & \mc{4}{c|}{{\textit{Top-level domain size in $x$-$y$: }$128\times128$}} &  \mc{2}{c}{$64\times64$} \\
  Grid size &$\omega$  & $^{lev=3, \alpha=0.2}_{coarse:\;2\times2\times 1}$  & $^{lev=3, \alpha=0.3^\dag}_{coarse:\;4\times 4 \times 1}$ & $^{lev=4, \alpha=0.4}_{coarse:\;2\times2\times 1}$ & $^{lev=4, \alpha=0.4}_{coarse:\;4\times4\times 1}$ &  $^{lev=3, \alpha=0.2}_{coarse:\;exact}$ &  $^{lev=4, \alpha=0.4}_{coarse:\;exact}$ \\
\midrule
$128\times128\times40$ &  1.36$\pi$  & 17 & 23 & 26 & 25  & 32 & 32 \\
$192\times192\times60^*$ &  2.02$\pi$ &22  & 31 & 36 & 36 & 33 & 43 \\
$256\times256\times80$ & 2.65$\pi$  & 41 & 46 & 56 & 57 & 79 & 99 \\
$384\times384\times120$ & 3.98$\pi$ & oom & oom  &  97 & 97 & oom & 332  \\
\bottomrule
 \end{tabular}
\caption{Three-dimensional experiments of the hybrid solver for the Overthrust model. $\dag$The attenuation was enlarged compared to 0.2 since the aggressive division of the coarsest grid solution led to inefficient solver. $^*$The top-level domain size is $192\times192$.}
\label{tab:hybridMGDD_3D_Overthrust}
\end{table}

\section{Conclusions and future work}

In this paper we present a new shifted Laplacian multigrid method for solving the elastic Helmholtz equation. Our main idea is to combine the shifted Laplacian with approaches for linear elasticity. The latter corresponds to the case of zero frequency in the elastic Helmholtz problem. With some specialized components in the multigrid cycle, such as mixed formulation and cell-wise relaxation, the elastic Helmholtz equation can be solved with the same efficiency as the acoustic Helmholtz equation only with respect to the shear wavenumber instead of the pressure wavenumber. We demonstrate this scaling property both in our experiments and our theoretical local Fourier analysis. 

To better handle realistic three-dimensional scenarios, where the system is huge, we harness the DD approach and combine it with multigrid. This way, we split the large problem into smaller problems and enhance parallelism of our solver. We use DD  both as a mean to distribute the fine-level problem, and as a solver for the coarsest level problem, that might still be large. We show that if we balance the degree in which the shifted Laplacian multigrid and DD simplify the problem (divisions in DD, coarsening in MG), we can have a solver whose performance is equivalent to either one of the methods alone. This balance is possible because of the local nature of both methods. Eventually we enjoy the computational advantages of both methods without harming each method's convergence by much. This combination makes our preconditioner highly applicable in 3D scenarios. 

Overall, this work demonstrates that besides the size, there is relatively little ``added difficulty'' to the elastic Helmholtz equation compared to the acoustic one, and that the two problems have a lot in common. The real question still remains: how to better deal with the indefiniteness of the Helmholtz problems (acoustic or elastic)---this is a main part of our future research. In addition, we will explore 
options to combine the multigrid approach with adaptive discretization, as the shear wavenumber tends to be much higher in the upper regions of the domain (requiring a finer mesh) than in the lower ones.
We will try combining different DD methods for the top-level and coarse grid DD, tuning the sweeping order and the ICs for each case.
In understanding the relationship between acoustic and elastic wave propagation problems, this work gives a valuable insight, which we hope that will lead to new elastic Helmholtz solvers based on acoustic ones.

%
\small
\bibliographystyle{elsarticle-num}
\bibliography{Helmholtzbib}

\end{document}